\def\equationautorefname~#1\null{Equation (#1)\null}
\def\sectionautorefname~#1\null{Section #1\null}
\def\subsectionautorefname~#1\null{Section #1\null}
\def\subsubsectionautorefname~#1\null{Section #1\null}
\def\figureautorefname~#1\null{Figure #1\null}
\documentclass[twocolumn]{aastex63}

\usepackage{amsmath}

\received{}
\revised{}
\accepted{\today}
\begin{document}

\title{Simulations of early kilonova emission from neutron star mergers} 

\correspondingauthor{Smaranika Banerjee}
\email{smaranikab@astr.tohoku.ac.jp}

%\author[0000-0002-0786-7307]{Smaranika Banerjee}
\author{Smaranika Banerjee}
\affiliation{Astronomical Institute, Tohoku University, Aoba, Sendai 980-8578, Japan}

\author{Masaomi Tanaka}
\affiliation{Astronomical Institute, Tohoku University, Aoba, Sendai 980-8578, Japan}

\author{Kyohei Kawaguchi}
%\affiliation{Institute for Cosmic Ray Research, The University of Tokyo, 5-1-5 Kashiwanoha, Kashiwa, Chiba 277-8582, Japan; Center for Gravitational Physics, Yukawa Institute for Theoretical Physics, Kyoto University, Kyoto, 606-8502, Japan}

\author{Daiji Kato}
\affiliation{National Institute for Fusion Science, National Institutes of Natural Sciences, Oroshi-cho, Toki, Gifu 509-5292, Japan}
\affiliation{Department of Advanced Energy Engineering Science, Kyushu University, Kasuga, Fukuoka 816-8580, Japan}

\author{Gediminas Gaigalas}
\affiliation{Institute of Theoretical Physics and Astronomy, Vilnius University, Saulėtekio av. 3, LT-10257 Vilnius, Lithuania}

\begin{abstract}
We present radiative transfer simulations for blue kilonovae hours after neutron star (NS) mergers by performing detailed opacity calculations for the first time.
We calculate atomic structures and opacities of highly ionized elements (up to the tenth ionization) with atomic number $Z = 20 - 56$. We find that the bound-bound transitions of heavy elements are the dominant source of the opacities in the early phase ($t < 1$ day after the merger),
and that the ions with a half-closed electron shell provide the highest contributions. The Planck mean opacity for lanthanide-free ejecta (with electron fraction of $Y_{\rm e}$ = 0.30 $-$ 0.40) can only reach around $\kappa \sim 0.5-1\,\rm{cm^{2}\,g^{-1}}$ at $t =$ 0.1 day, whereas that increases up to $\kappa \sim 5 - 10\,\rm{cm^{2}\,g^{-1}}$  at $t =$ 1 day.
The spherical ejecta model with an ejecta mass of $M_{\rm ej} = 0.05M_{\odot}$
gives the bolometric luminosity of $\sim\,2\times\, 10^{42}\,\rm erg\,s^{-1}$ at $t\,\sim$ 0.1 day.
We confirm that the existing bolometric and multi-color data of GW170817 can be naturally explained by the purely radioactive model.
The expected early UV signals reach 20.5 mag at $t \,\sim$ 4.3 hours for sources even at 200 Mpc, which is detectable by the facilities such as Swift and the Ultraviolet Transient Astronomy Satellite (ULTRASAT). 
The early-phase luminosity is sensitive to the structure of the outer ejecta, as also pointed out by \citet{kasen2017}. 
Therefore, the early UV observations give strong constraints on the structure of the outer ejecta as well as the presence of a heating source besides $r$-process nuclei.

\end{abstract}


\section{Introduction} \label{sec:intro}
Compact object mergers (neutron star-neutron star (NS-NS) or neutron star-black hole mergers) have long been hypothesized to be the sites for heavy element synthesis \citep{Lattimer1974,Eichler1989,Freiburghaus1999, Korobkin2012,Wanajo2014}. In the material ejected from the mergers, a rapid neutron-capture nucleosynthesis ($r$-process) takes place. Radioactive decay of heavy elements can give rise to electromagnetic transients in the ultraviolet, optical, and near infrared (UVOIR) spectrum, similar to supernovae \citep{Li1998, Kulkarni2005} but on a faster timescale ($\sim 1-7$ days) and with lower peak luminosities \citep{Metzger2010,Roberts2011,kasen2013,tanaka2013}. These transient are called kilonovae \citep{Metzger2010} or macronovae \citep{Kulkarni2005}.  

These compact object mergers are also the source of gravitational waves (GWs) in the LIGO/Virgo detection frequency range, making them ideal targets for multi-messenger observation. In fact, the first detection of an NS-NS merger event, GW170817 \citep{Abbott2017a}, was accompanied by emissions in the wide range of the electromagnetic spectrum. The coincident detection of a short gamma-ray burst (GRB) at $t\,\sim$ 2 s (where $t$ is the time since the merger) proved the association between short GRBs and NS merger events \citep{Connaughton2017, Savchenko2017}. The optical and near-infrared emissions were detected at $t\sim$ 11 hours \citep{Coulter2017, Yang2017,Valenti2017}, followed by the detection of a bright UV emission by Swift \citep{Evans2017} at $t\sim$ 16 hours. X-ray and radio afterglow were also detected at $t =$ 9 days and $t =$ 16 days, respectively \citep{Troja2017,Hallinan2017,Mooley2017}. This extensive dataset for GW170817 provides us with a novel way to probe various physical aspects of NS mergers.

In this work, we focus on emissions in the UVOIR spectrum.
The fast decline of the light curve in the optical band and late-time brightening in the near infrared (NIR) band are well explained by kilonova \citep{kasen2017,tanaka2017,Shibata2017,Perego2017,Rosswog2018,Kawaguchi2018}. However, the origin of the bright UV and blue emissions in the early time ($t<1$ day) is not yet clear \citep{arcavi2018}.
This early-time behavior could be explained by the kilonova, as in the later phase.
In fact, one-component model by \citet{Waxman2018} and multi-component model by \citet{Villar2017} give reasonable agreement with the early phase data.
Alternatively, the early emission may be the result of emission from the ejecta heated by the cocoon, formed by the interaction of the relativistic jet with the surrounding ejecta \citep{mansi2017,Piro2018}.
Other possibilities include emission powered by $\beta$-decays of free neutrons \citep{Metzger2015, Gottlieb2020}
or by a long-lived central engine \citep{Metzger2008, Yu2013, Metzger2014, Matsumoto2018,Metzger2018,Li2018,Wollaeger2019}.

One of the uncertainties all the models share is lack of atomic data at early times. A few hours after the merger, the ejecta are still hot ($T\,\sim\,10^5$ K), with $r$-process elements in the ejecta highly ionized. However, there was no atomic data for such conditions and subsequently no data for the opacity. Previous works have used different strategies to tackle this problem; for example, \citet{Waxman2018} considered a functional form for the time-dependent opacity. 
However, \citet{Villar2017} used a fixed value of opacity for different segments of the ejecta in their multi-component ejecta model. Similarly, in the models of cocoon emission and free neutron decay \citep{mansi2017,Piro2018,Gottlieb2020}, the opacity was fixed at a certain value.

In fact, there have been several efforts to evaluate the opacity from atomic models.
The earliest works attempted the calculation for only a few representative elements \citep{kasen2013, tanaka2013, Fontes2017, Wollaeger2017, tanaka2018}, assuming that the overall ejecta opacity can be reflected by these elements. More recently, atomic opacity data for all lanthanides ($Z$ = 58 $-$ 70, \citealt{kasen2017, Fontes2020}) and finally all the $r$-process elements ($Z$ = 31 $-$ 88, \citealt{Tanaka2020a}) have been calculated. However, these works considered the maximum ionization to be the fourth or third degree, which is only a reasonable assumption for the condition of the ejecta around $t \sim$ 1 day.

In this paper, we perform the first opacity calculation for the highly ionized light $r$-process elements ($Z$ = 20 $-$ 56), suitable for describing the ejecta condition as early as hours after a compact object merger. Estimates of different opacity components, excluding bound-bound opacity, are shown in \autoref{sec:opacity}. Calculations of the bound-bound opacity from atomic structure models are separately discussed in \autoref{subsec:bb}. In \autoref{sec:radtransfer}, we perform radiative transfer simulations with the newly calculated opacity. The application of our results to the early-time data of GW170817, as well as the future prospects, are discussed in \autoref{sec:discussion}. Finally we provide concluding remarks in \autoref{sec:conclusion}. Throughout the paper AB magnitude system is used.

\section{Opacities at early time} \label{sec:opacity}

In this section, we discuss the behaviors of different opacity components in NS merger ejecta. Different processes including electron scattering, free-free transitions,  bound-free (or photo-ionization) transitions, and bound-bound transitions contribute to the total opacity.
In earlier works of supernova \citep{Pinto2000} and kilonova \citep{kasen2013, tanaka2013}, it is found that the main contribution to the opacity comes from the bound-bound transitions. Since our work focuses on an early phase, we reevaluate the contribution from each of the opacity components. 
 
 A few hours after the merger, the ejecta are dense ($\rho\,\sim\,10^{-10}\,\rm{g\,cm^{-3}}$) and hot ($T\,\sim \,10^{5}$ K). Heavy elements in the ejecta are highly ionized under such conditions. 
 By solving the Saha ionization equation, under the assumption of local thermal equilibrium (LTE) for single-element ejecta, we find that the ionization of the elements reach at least tenth degree (XI in spectroscopic notation) at $T\,\sim 10^{5}$ K for $\rho\,\sim\,10^{-10}\,\rm{g\,cm^{-3}}$. As the temperature at which the ionization reaches the tenth degree varies not so significantly for different elements, we carry out our analysis considering the maximum ionization fixed to the tenth degree (XI) for the rest of the paper.

The primal goal of our present study is to calculate kilonova light curves in the early phase ($t < 1$ day). As the early light curves of GW170817 and AT2017gfo are interpreted as so-called "blue" kilonova, with a small fraction or no lanthanide elements \citep{Metzger2010,Roberts2011,Fernandez2014}, we focus on the elements with atomic number of $Z = 20 - 56$ to calculate different opacity components.

In the following subsections, we discuss different opacity components in the early phase. The bound-bound opacity is discussed in \autoref{subsec:bb} as it requires extensive atomic calculations.
Although the NS merger ejecta consist of a mixture of elements, we first discuss the opacity for single-element ejecta to analytically estimate the opacities at early times. 
We consider the mixture of elements for the calculations of the bound-bound opacities in \autoref{subsubsec:mix_bbop},
and consider all the opacity components in our final radiative transfer simulations in \autoref{sec:radtransfer}.

\subsection{Electron scattering opacity} \label{subsec:es}
As the ionization is high in the early phase, electron scattering can conceivably play an important role in the opacity. 
The number density ($n_{\rm{e}}$) of the electrons in the single-element ejecta can be estimated as
\begin{equation}\label{eqn:ne}
    n_{\rm{e}}  = \frac{\rho}{Am_{\rm{p}}}j,
\end{equation}
where $A$ is the mass number, $m_{\rm{p}}$ is the mass of the proton, and $j$ is the ionization degree ($j$ = 10 for tenth (XI) ionization) of an element.
From this, the electron scattering opacity ($\kappa^{\rm{es}}$) can be calculated via
\begin{equation}\label{eqn:es}
    \kappa^{\rm{es}} = \frac{n_{\rm{e}}\sigma_{\rm{Th}}}{\rho} = \frac{\sigma_{\rm{Th}}}{Am_{\rm{p}}}j,
\end{equation} 
where $\sigma_{\rm{Th}}$ is the Thomson scattering cross section.
 For the single-element ejecta with maximum ionization, i.e. $j$ = 10, the electron scattering opacity is estimated as $\kappa^{\rm{es}}\,=\,(3\,-10)\,\times\,10^{-2} \,\rm{cm^2\,g^{-1}}$ for elements $Z$ = 20 $-$ 56, with a greater opacity value for the lower $Z$ (and thus lower $A$) elements. For iron (Fe, $Z$ = 26), the value is $7\times10^{-2}\, \rm{cm^2\,g^{-1}}$.

\begin{figure}[t]
\includegraphics[width=\linewidth]{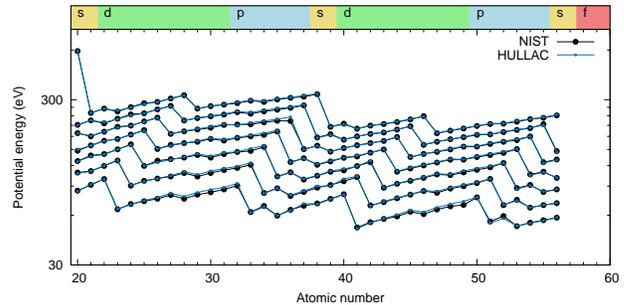}
\caption{Comparison of the ionization potentials calculated with the Hebrew University Lawrence Livermore Atomic Code (HULLAC, blue), with the potentials listed in the National Institute of Science and Technology atomic spectra database (NIST ASD, black, \citealt{kramida18}). Ionization potentials from ion V to XI are shown from the bottom image to the top image, respectively. The colors in the top panel are characterized by the outermost electron shells for singly ionized ions, as in previous works \citep{kasen2013,tanaka2018,Tanaka2020a}. \label{fig:epot}}
\end{figure}
\subsection{Free-free opacity}\label{subsec:ff}
Free-free transitions constitute another component of the total opacity ($\kappa^{\rm{ff}}_{i,j}(\lambda)$). For a particular element $i$ at an ionization state $j$, this opacity component can be calculated as in \citet{Rybicki1986}:
\begin{equation}\label{eqn:eqn_ff}
\begin{split}
    \kappa^{\rm{ff}}_{i,j}(\lambda) = \frac{4e^{6}\lambda^{3}}{3 m_{\rm{e}}hc^{4}\rho}\bigg(\frac{\rm{2}\pi}{3km_{\rm{p}}}\bigg)^{\frac{1}{2}}T_{\rm e}^{-\frac{1}{2}}(j-1)^2 n_{\rm{e}}n_{i,j} \\ \times(1 - e^{-\frac{hc}{\lambda kT_{\rm e}}})\bar{g}_{\rm{ff}},
\end{split}
\end{equation}
where $T_{\rm e}$ is the electron temperature (for which we substitute the common temperature, $T$, under LTE), $\overline{g}_{\rm{ff}}$ is the velocity-averaged free-free Gaunt factor (which is fixed at unity, following the method of \citealt{tanaka2013}), and $n_{i,j}$ is the ion density, estimated as
\begin{equation}\label{eqn:ion_fr}
    n_{i,j} = \frac{f_{i,j}X_{i}\rho}{Am_{\rm{p}}}
\end{equation}
Here $X_{i}$ is the fraction of $i$th element in the ejecta and $f_{i,j}$ is the fraction of $i$th element at a $j$th ionization state.
To obtain an analytic estimate of the free-free opacity component for single-element ejecta, the electron density is calculated from \autoref{eqn:ne}, while the ion density is estimated by putting $f_{i,j}\,=1$ and $X_{i}\,=1$ in \autoref{eqn:ion_fr}. We find that a single-element ejecta, with temperature $T=\,10^{5}$ K and density $\rho =10^{-10}\,\rm g\, cm^{-3}$, has a free-free opacity component of $\kappa^{\rm{ff}}_{i,j}= (2-3) \times10^{-4}\,\rm{cm^2\,g^{-1}}$ at a wavelength $\lambda$ = 1000 $\rm\AA$ for $Z$ = 20 $-$ 56. This opacity component is greater for lower $Z$ elements. For Fe ($Z$ = 26), $\kappa^{\rm{ff}}_{i,j}\,=\,2.6\times10^{-4}\,\rm{cm^2\,g^{-1}}$. Thus, even in the early phase, the free-free transition opacity component is relatively small.

\subsection{Bound-free opacity}\label{subsec:bf}
Another process contributing to the opacity is photo-ionization or bound-free transition. The bound-free transition opacity is calculated by 
\begin{equation}\label{eqn:eqn_bf}
    \kappa^{\rm{bf}}_{i,j}(\lambda) = \frac{n_{i,j} \sigma^{\rm{bf}}_{i,j}}{\rho} ,
\end{equation}
where $\sigma^{\rm{bf}}_{i,j}$ is the bound-free cross section for the $i$th element in the $j$th ionization state. The bound-free cross section is estimated from a fitting formula taken from \citet{Verner1996}. For Fe ($Z$ = 26) in the tenth (XI) ionization state for $T\,=\,10^{5}$ K, the cross section is $\sigma^{\rm{bf}}_{i,j}\,=\,0.45$ Mb at the ionization threshold.

The elements with atomic numbers $Z$ = 20 $-$ 56 have a tenth ionization potential energy $\geq$ 250 eV (\autoref{fig:epot}), corresponding to a wavelength of $\lambda\,\leq$ 50 \AA. According to the blackbody function at a temperature of $T = 10^{5}$ K, the fraction of photon energy present at such a short wavelength range is $\sim\,10^{-6}$.
Calculating the same for different ionization states of different elements in a temperature range of $T\,=\,10^{3}\,-\,10^{5}$ K, we find that the fraction never reaches beyond $10^{-4}$.
Therefore, although the photo-ionization cross-section itself is high, the number of photons with energy greater than the ionization potential is negligible. Therefore, bound-free opacity component does not significantly contribute to the total opacity. 

There is no available bound-free cross section data for the elements with $Z\,>$ 26, i.e., Fe. Following the method adopted by \citet{tanaka2013}, we use the cross sections of Fe for elements with a higher $Z$ in the radiative transfer simulation (\autoref{sec:radtransfer}). This crude approximation does not alter the results since the bound-free transition opacity is not predominant, as discussed above.

\begin{figure*}[t]
\begin{center}
  \begin{tabular}{c}
    % 1 (first figure)
     \begin{minipage}{0.5\hsize}
      \begin{center}
        \includegraphics[width=\linewidth]{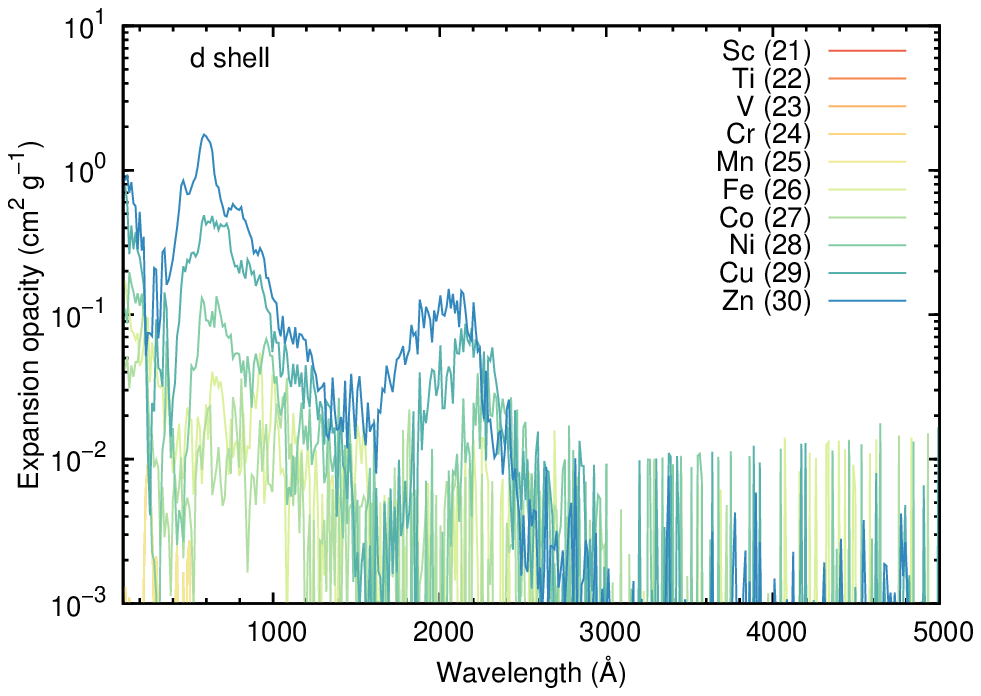}
      \end{center}
    \end{minipage}
    
    \begin{minipage}{0.5\hsize}
      \begin{center}
        \includegraphics[width=\linewidth]{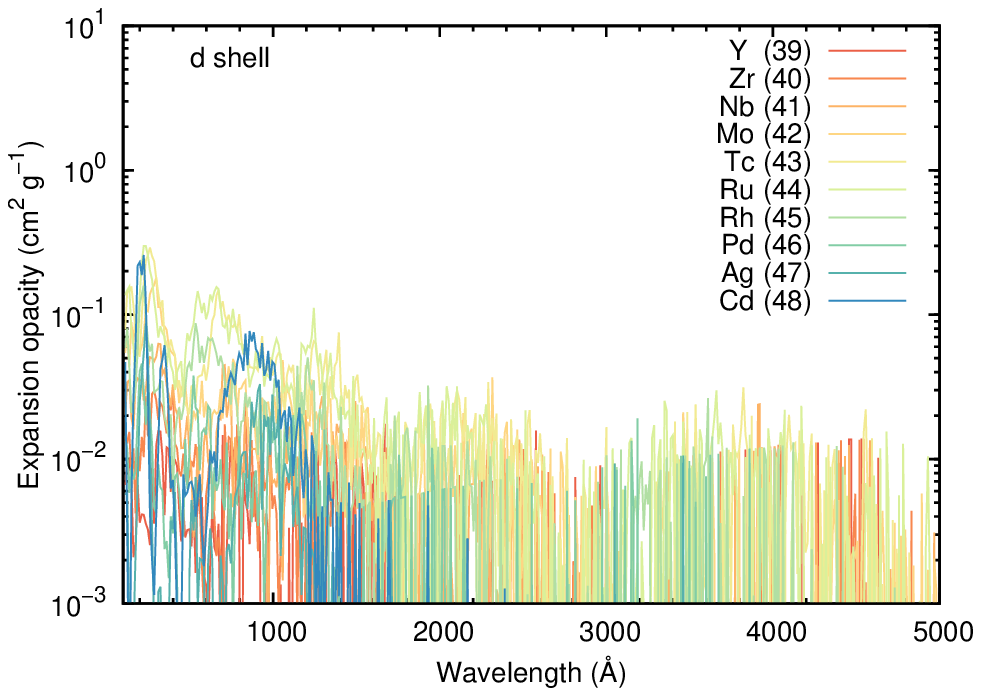}
      \end{center}
    \end{minipage}
    
    \\
   
   \begin{minipage}{0.5\hsize}
      \begin{center}
        \includegraphics[width=\linewidth]{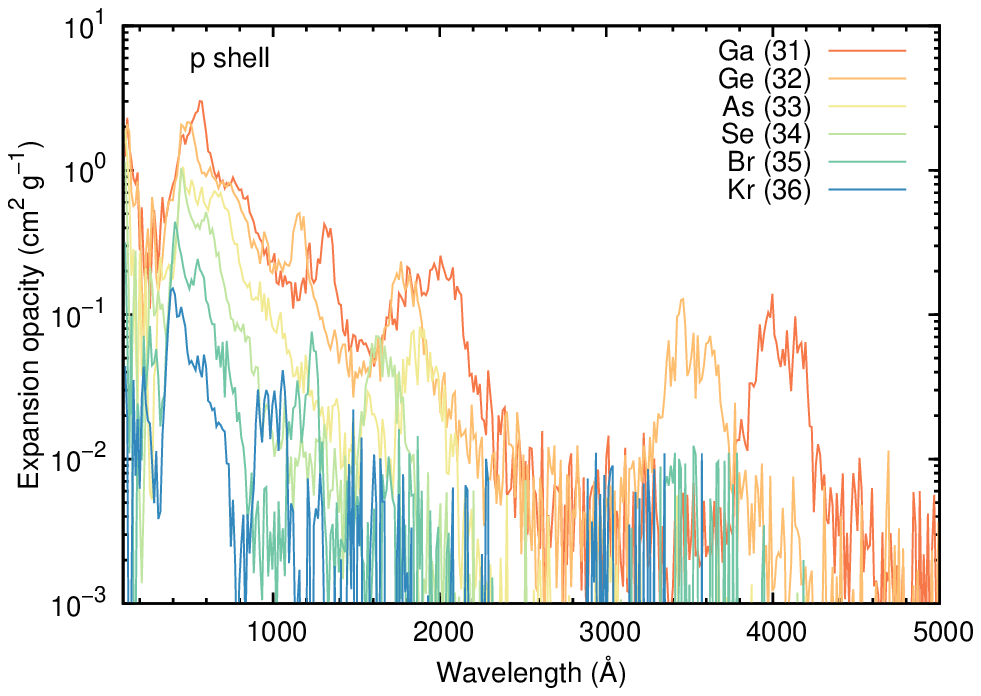}
      \end{center}
    \end{minipage}
    
     \begin{minipage}{0.5\hsize}
      \begin{center}
        \includegraphics[width=\linewidth]{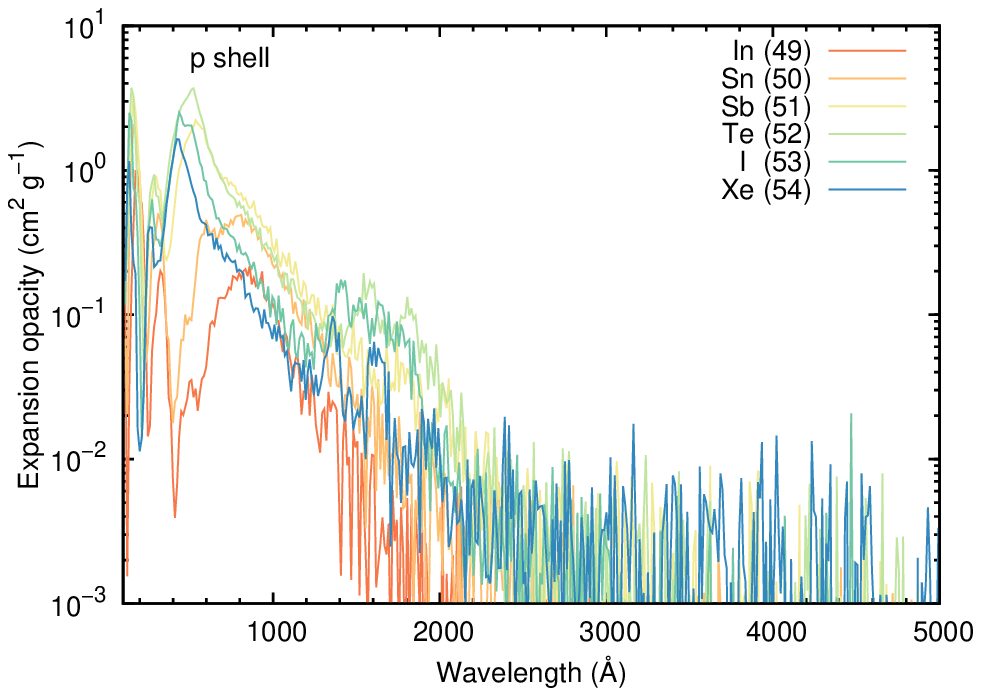}
      \end{center}
    \end{minipage}
    \\
    \begin{minipage}{0.5\hsize}
      \begin{center}
        \includegraphics[width=\linewidth]{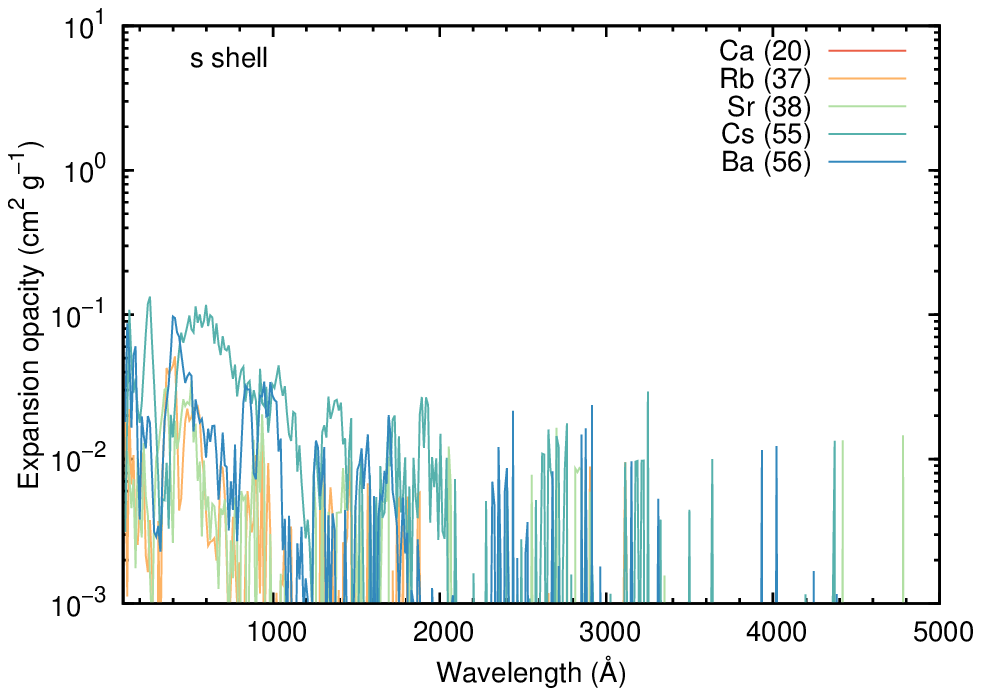}
      \end{center}
    \end{minipage}
  \end{tabular}
  \caption{The expansion opacity as a function of wavelength at $T\,=\,10^{5}$ K, $\rho \,=\, 10^{-10}\, \rm g\,cm^{-3}$, and $t = 0.1$ day for \textbf{Top:} $d$-shell elements, \textbf{Middle:} $p$-shell elements, \textbf{Bottom:} $s$-shell elements.}
  \label{fig:op_wav}
  \end{center}
\end{figure*}
% \newline

\section{Bound-bound opacity}\label{subsec:bb}

To evaluate the bound-bound opacity for the NS merger ejecta, we require extensive data on energy levels and transition probabilities for heavy elements. Since complete data calibrated with experiments are not available, we first perform the atomic structure calculations to construct the line list in \autoref{subsubsec:at_struc}. Using those results, we evaluate the bound-bound opacities in \autoref{subsubsec:bb_op}.

\subsection{Calculations of atomic structure}\label{subsubsec:at_struc}

We perform atomic structure calculations by using the Hebrew University Lawrence Livermore Atomic Code (HULLAC) \citealt{bar-shalom01}). The calculation methods follow those adopted by \citet{Tanaka2020a}, where the calculation was limited from neutral atoms (I) to triply ionized ions (IV). We extend the calculations up to the tenth ionization state (XI) for elements with $Z$ = 20 $-$ 56.

Past atomic calculations of $r$-process elements for kilonovae ejecta could only achieve typical energy level accuracies of a few tens of percent \citep{kasen2013,tanaka2018}. This is not particularly accurate compared to standard accuracy measurements in atomic physics. Complete and accurate calculations for $r$-process elements with several excited levels are still difficult to achieve \citep{gaigalas19,radziute20}.
The inaccuracy in the atomic calculations typically result in a systematic uncertainty in the bound-bound opacity by a factor of around $\sim$ 2 \citep{kasen2013,gaigalas19}.
As discussed below, evaluating the accuracy for highly ionized ions is not currently possible. 
Therefore, we assert that the factor of 2 uncertainty also exists in the opacities given in this paper.

The main difference in our calculations compared to \citet{Tanaka2020a} is the configurations included in the atomic calculations. For the highly ionized ions considered in this paper, information on energy levels and electronic configurations is lacking. Therefore, when only a few configurations are listed in the National Institute of Science and Technology atomic spectra database (NIST ASD, \citealt{kramida18}), we implement the configurations of isoelectronic neutral atoms. A typical number of included configurations for each ion is 13.
This assumption provides some convergence in the opacity, typically within an uncertainty of less than 10\% \citep{Tanaka2020a}; a small enough value compared to the expected systematic uncertainty.

Since the available data for energy levels in the NIST ASD are limited, we are unable to evaluate the accuracy of our calculations with well-evaluated data. Instead, we compare the calculated ionization potentials with those in the NIST ASD (\autoref{fig:epot}). The mean accuracy is found to be 1.6\%, 1.1\%, 1.0\%, 0.8\%, 0.8\%, 0.6\%, and 0.6\% for ion V - XI, respectively. Highly ionized ions have a fewer number of bound electrons and thus the system becomes simpler. Also correlation converges more rapidly when ionization of elements increases. As a result, the accuracy for the highly ionized ions are much better than that obtained for lower ionization states (4--14\%, \citealt{Tanaka2020a}).

\subsection{Calculations of bound-bound opacity}\label{subsubsec:bb_op}

Equipped with the atomic data for highly ionized ions, we calculate the bound-bound opacity for early-time ejecta. In supernovae and NS mergers, the matter is expanding with a high velocity and high velocity gradient. In such a system, the opacity can be enhanced \citep{Karp1977}. To calculate this opacity, we use the expansion opacity formalism \citep{Eastman1993}:
\begin{equation}\label{eqn:kexp}
    \kappa_{\rm{exp}}(\lambda) = \frac{1}{ct\rho}\sum_{l}\frac{\lambda_{l}}{\Delta \lambda}(1 -e^{-\tau_{l}}),
\end{equation}
where $\lambda_{l}$ is the transition wavelength in a wavelength interval $\Delta \lambda$, and $\tau_{l}$ is the Sobolev optical depth at the transition wavelength, calculated as
\begin{equation}\label{eqn:tau}
    \tau_{l} = \frac{\pi e^{2}}{m_{\rm{e}} c} n_{i,j}\lambda_{l}f_{l}t\frac{g_{l}}{g_{0}}e^{-\frac{E_{l}}{kT}}.
\end{equation}
Here $E_{l}$, $g_{l}$, and $f_{l}$ are the energy, statistical weight of the lower level of the transition, and strength of transition, respectively. The statistical weight of the ground state is expressed as $g_{0}$. 

\begin{figure*}[t]
% \begin{center}
  \begin{tabular}{c}
    %  \centering
    \begin{minipage}{0.5\hsize}
      \begin{center}
        \includegraphics[height=6cm,width=8cm,keepaspectratio]{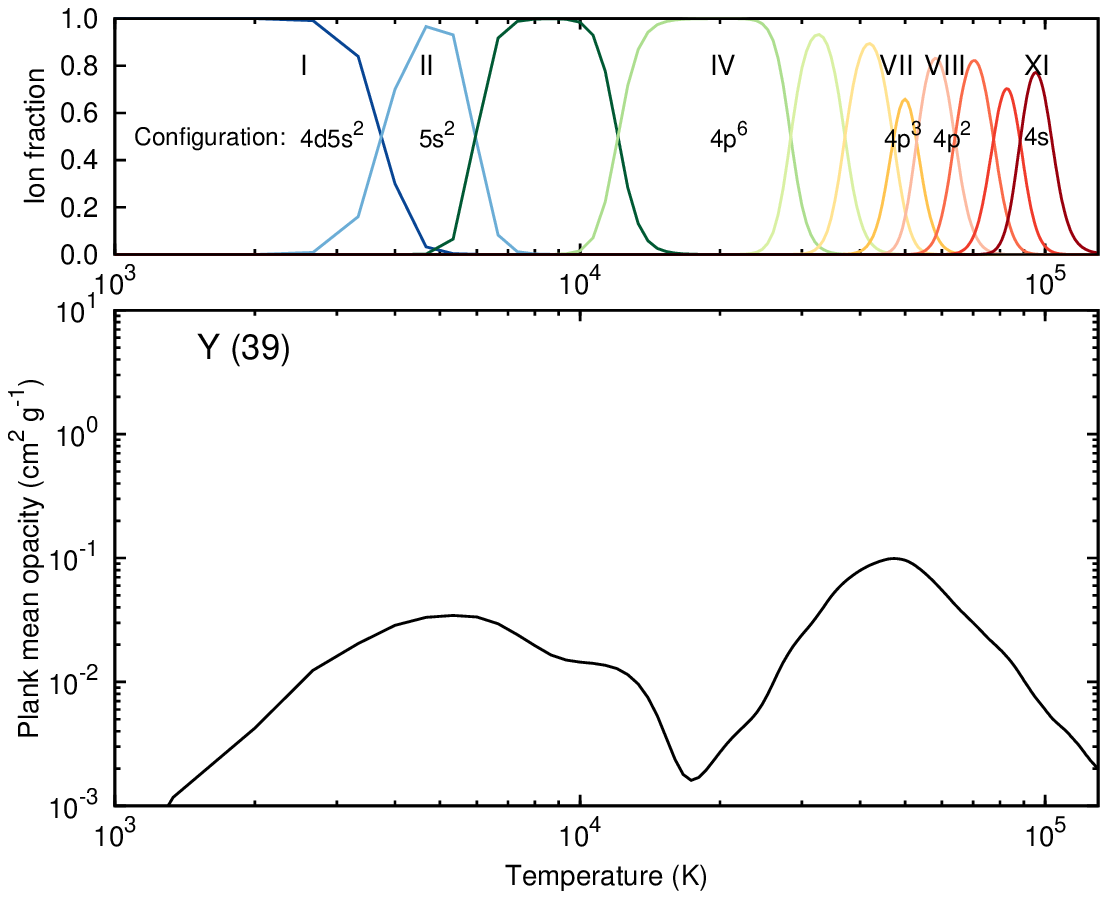}
      \end{center}
    \end{minipage}
    
    \begin{minipage}{0.5\hsize}
      \begin{center}
        \includegraphics[height=6cm,width=8cm,keepaspectratio]{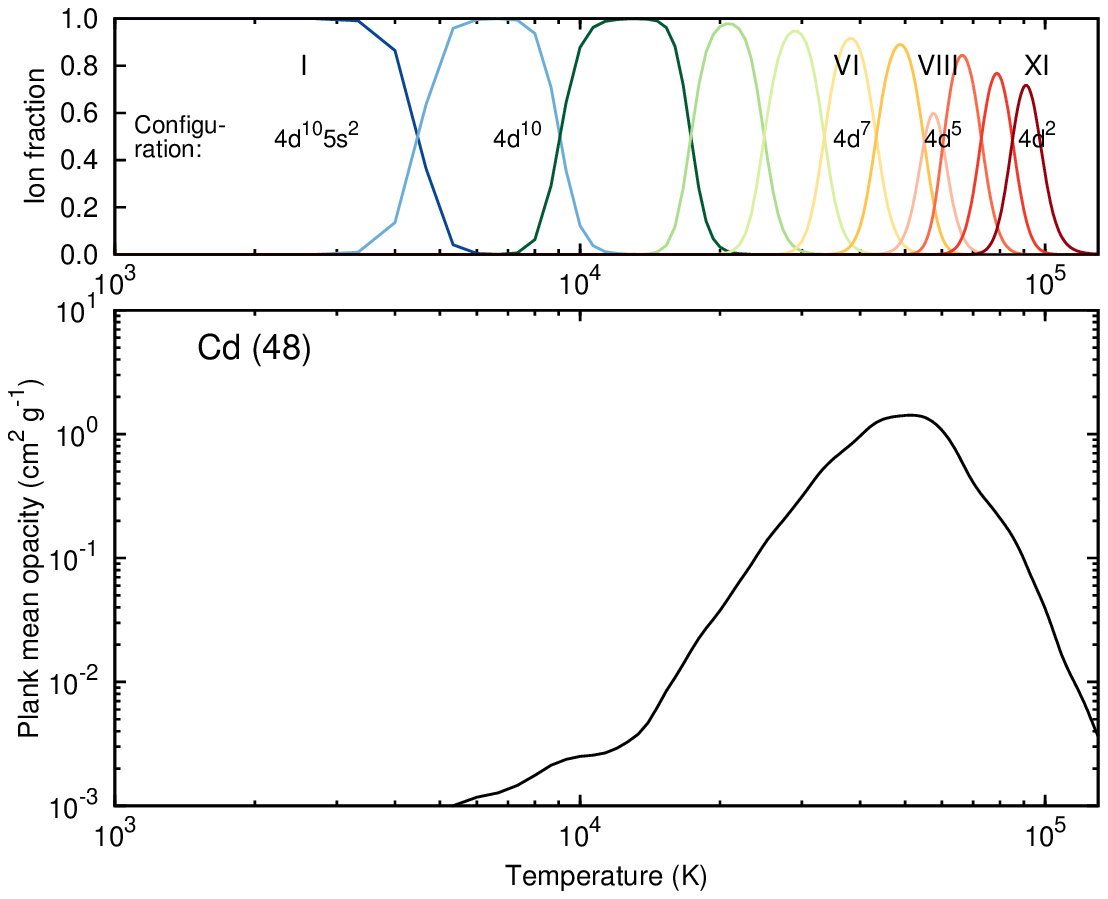}
      \end{center}
    \end{minipage}
          \\

     \begin{minipage}{0.5\hsize}
      \begin{center}
        \includegraphics[height=6cm,width=8cm,keepaspectratio]{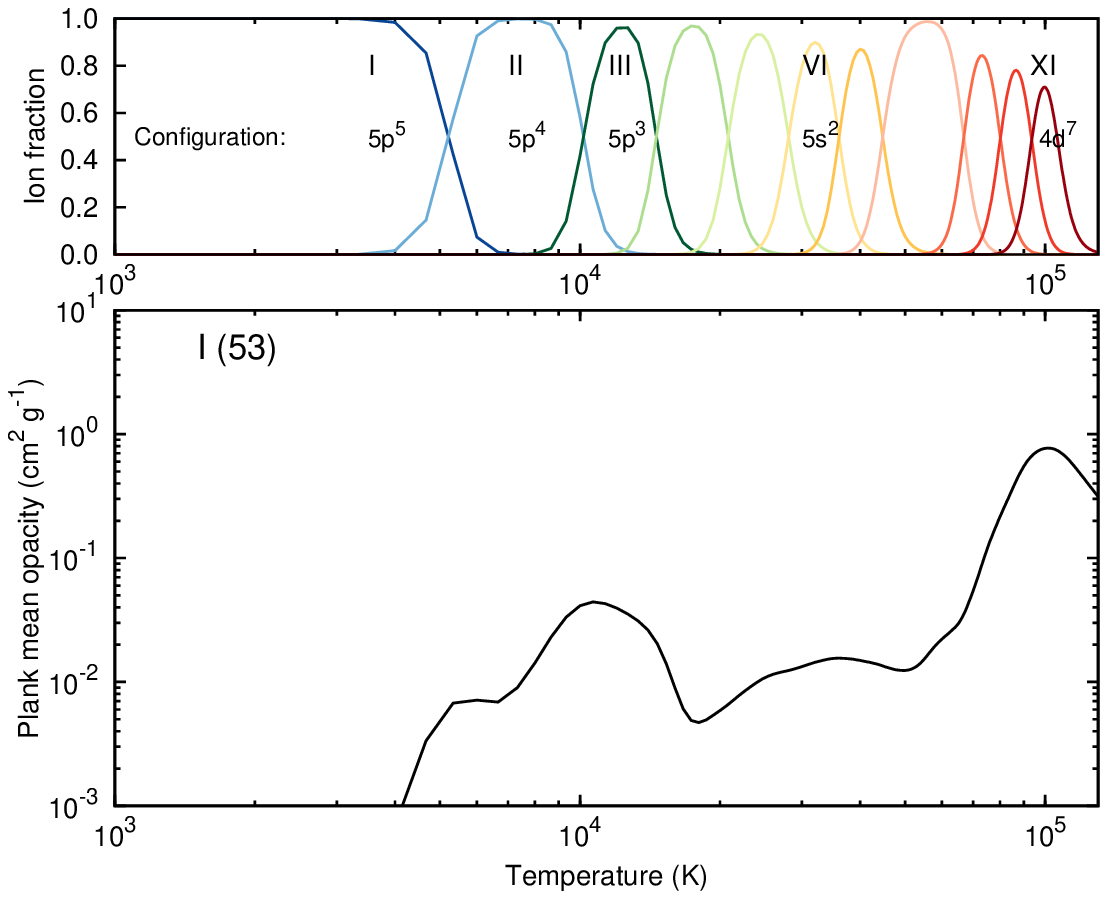}
      \end{center}
    \end{minipage}
         
    \begin{minipage}{0.5\hsize}
      \begin{center}
        \includegraphics[height=6cm,width=8cm,keepaspectratio]{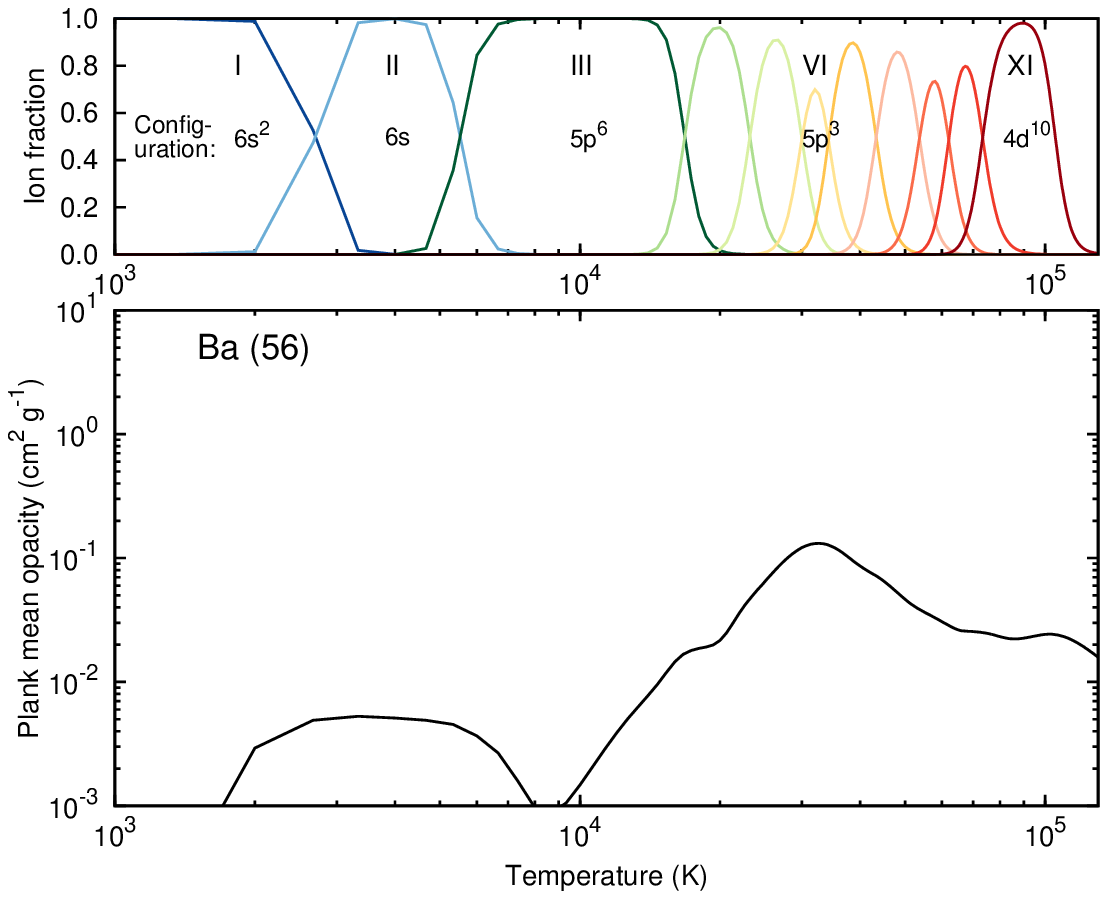}
      \end{center}
    \end{minipage}
  \end{tabular}
  \caption{\textbf{Top left:} The upper panel shows the how ion fraction changes with the temperature for $d$-shell element Y ($Z$ = 39). The effective shell structures for the ionization states of Y, where opacity varies significantly, are also shown. The bottom panel shows the variation of the Planck mean opacity with temperature. \textbf{Top right:} The same for $d$-shell element Cd ($Z$ = 48). \textbf{Bottom left:} The same for $p$-shell element I ($Z$ = 53). \textbf{Bottom right:} The same for $s$-shell element Ba ($Z$ = 56).}
  \label{fig:op_ion_Pl}
%   \end{center}
\end{figure*}

\subsubsection{Opacity of individual element}\label{subsubsec:elem_bbop}
We calculate the expansion opacity as a function of wavelength for the elements with $Z\, =\, 20 - 56$. The elements are categorized as either $d$, $p$, or $s$-shell elements according to the electron configurations of their neutral state (in \autoref{fig:op_wav}, $d$, $p$, and $s$-shell elements are shown in the top, middle, and bottom panels, respectively). The temperature and density are assumed to be $T\,\sim \,10^{5}$ K and $\rho\,=\,10^{-10} \,\rm g\,cm^{-3}$; typical conditions at $t \,=$ 0.1 day. Depending on the element, the expansion opacity varies, with $\kappa_{\rm exp}\,= 0.001 - 4 \rm \,cm^{2}\, g^{-1}$. The opacity is higher at UV wavelengths, similar to the behavior at later time .

The temperature dependence of the expansion opacity can be understood by calculating the Planck mean opacity, $\kappa_{\rm{mean}}$. 
\begin{figure*}[t]
\begin{center}
  \begin{tabular}{c}
     
 \begin{minipage}{0.5\hsize}
      \begin{center}
        \includegraphics[width=\linewidth]{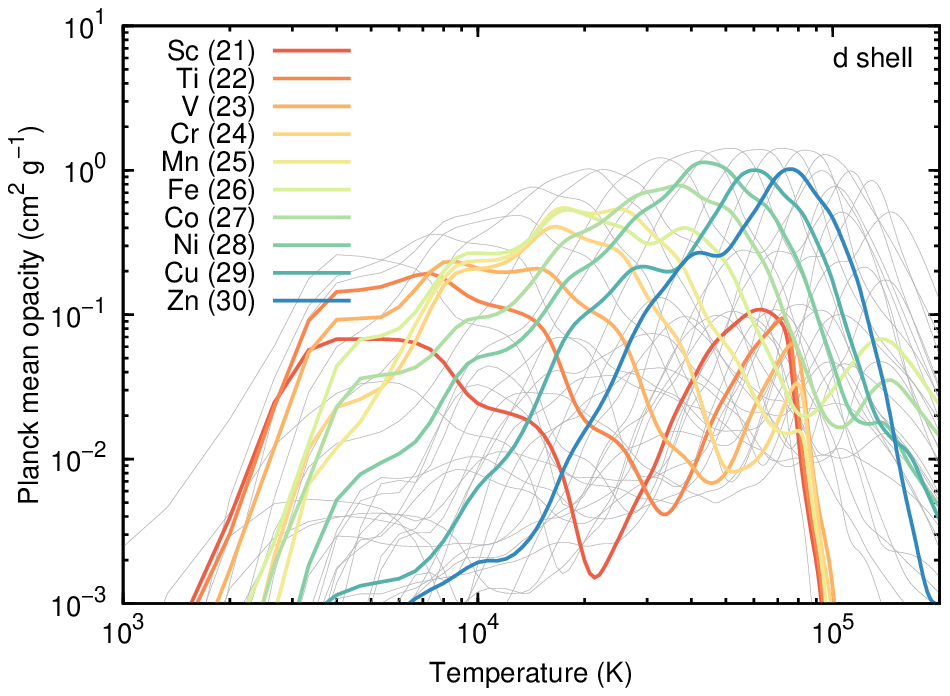}
      \end{center}
    \end{minipage}
 
     \begin{minipage}{0.5\hsize}
      \begin{center}
        \includegraphics[width=\linewidth]{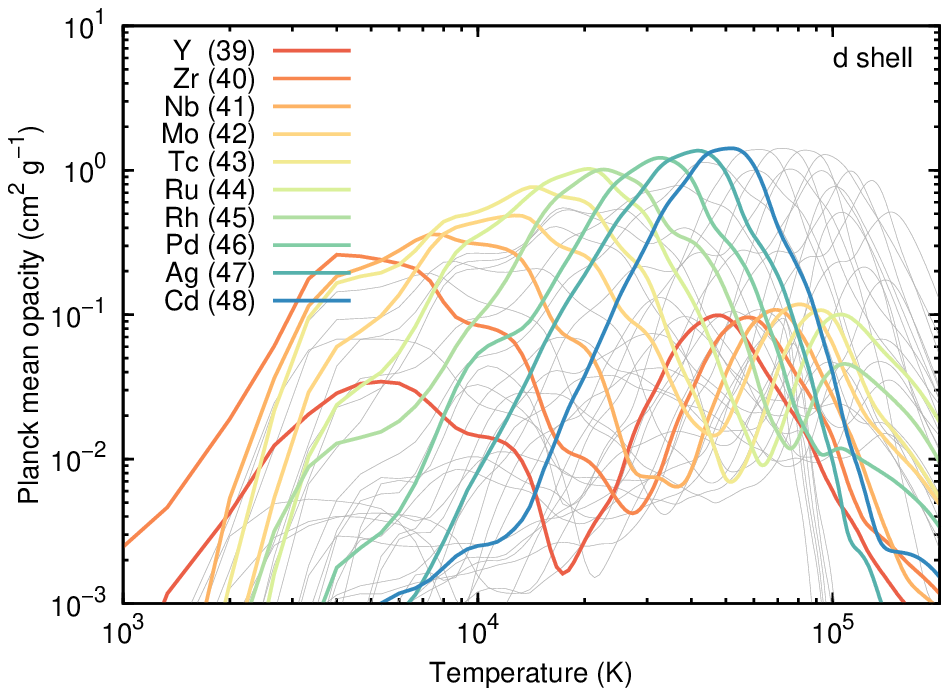}
      \end{center}
    \end{minipage}
    \\
    \begin{minipage}{0.5\hsize}
      \begin{center}
        \includegraphics[width=\linewidth]{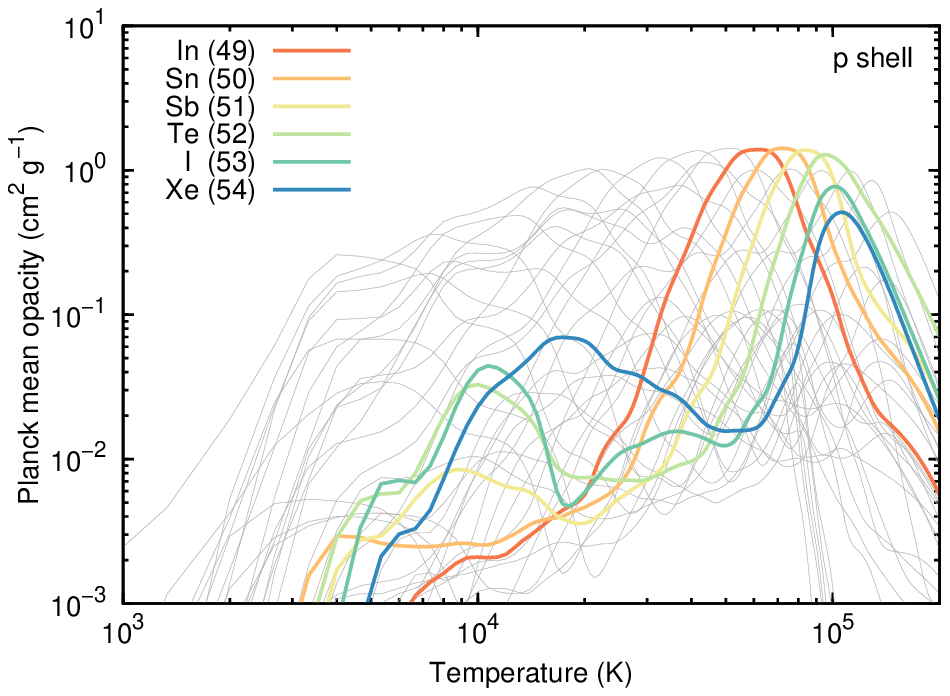}
      \end{center}
    \end{minipage}
    
     \begin{minipage}{0.5\hsize}
      \begin{center}
        \includegraphics[width=\linewidth]{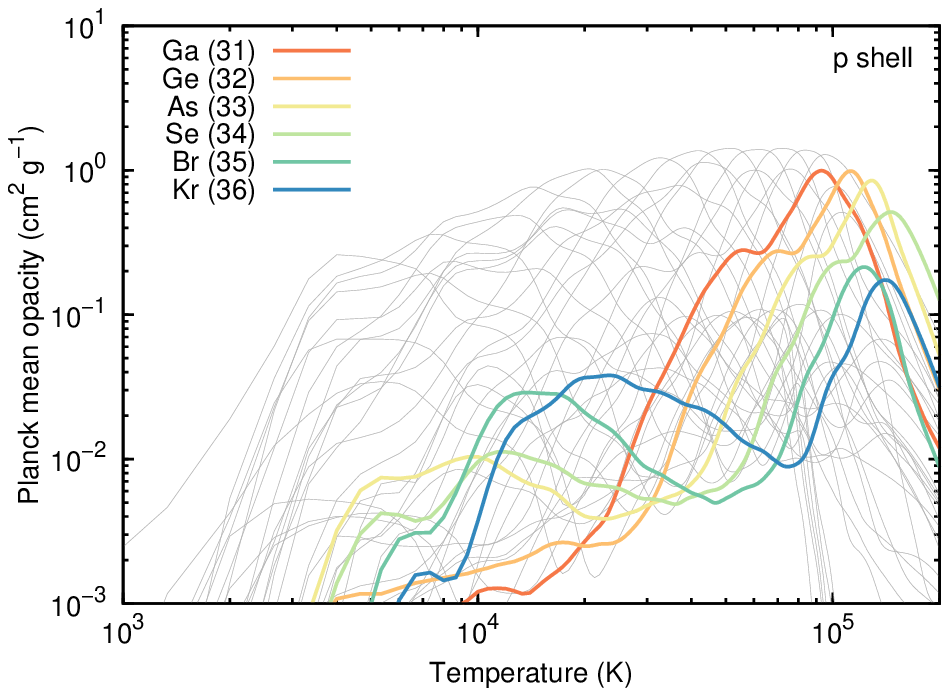}
      \end{center}
    \end{minipage}
    \\
    \begin{minipage}{0.5\hsize}
      \begin{center}
        \includegraphics[width=\linewidth]{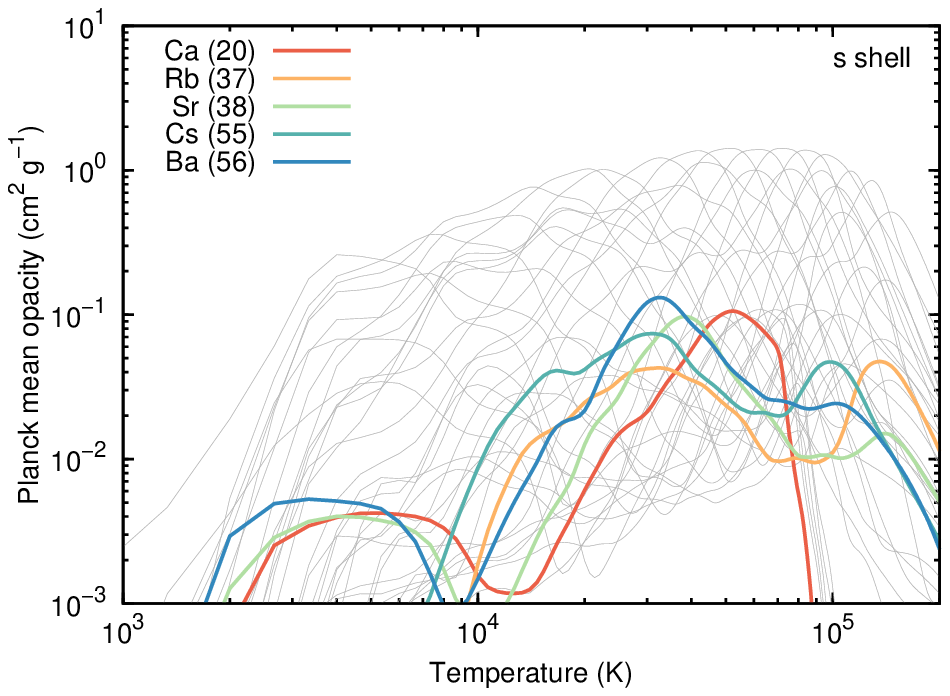}
      \end{center}
    \end{minipage}
  \end{tabular}
  \caption{\textbf{Top:} The Planck mean opacity as a function of temperature for $d$-shell elements at $\rho\, =\, 10^{-10} \,\rm g\,cm^{-3}$ and $t = 0.1$ day. \textbf{Middle:} The same for $p$-shell elements. \textbf{Bottom:} The same for $s$-shell elements.}
  \label{fig:op_Pl}
  \end{center}
\end{figure*}

\begin{figure*}[t]
\begin{center}
\includegraphics[width=0.6\textwidth]{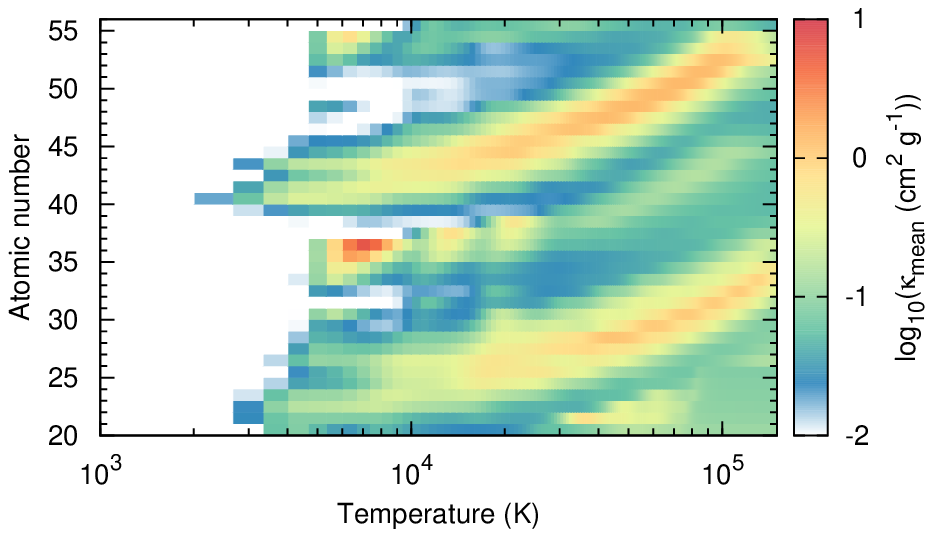}
\caption{The Planck mean opacity for different elements and temperature. The ejecta conditions are $t =$ 0.1 day and $\rho\, = \,10^{-10}\, \rm g\,cm^{-3}$.}\label{fig:ZvsTK}
\end{center}
\end{figure*}
Since the overall variation of the mean opacity is different for each element, we first discuss the trend for a few representative elements from different shells (\autoref{fig:op_ion_Pl}). There are two main factors that determine the trend of opacity for highly ionized ions: half-closed shells have the highest complexity measure, and the increment in $Z$ in a shell raises the energy distribution upwards \citep{Tanaka2020a}. 
The Boltzmann statistics predicts that lower energy levels are more populated and the transitions from such levels contribute to the opacity the most. 
At moderate temperatures, the elements or ions with half-closed shells do not necessarily have the highest opacity because their energy levels are pushed toward higher energies. At higher temperatures, higher energy levels are more populated and the opacities of the ions with half-closed shell are greater than other elements
within the same shell.

As the ionization degree of $d$-shell element Yttrium (Y, $Z$ = 39) increases with temperature, the Planck mean opacity evolves as shown in top left panel of \autoref{fig:op_ion_Pl}. When Y is singly or doubly ionized (II $-$ III) at $T = 5000 - 10000$ K, it has a similar energy level distribution to the neutral $s$-shell elements Strontium (Sr, $Z$ = 38) and Rubidium (Rb, $Z$ = 37). These elements contain only a few strong transitions. When Y becomes triply ionized (IV), it has a closed $p$-shell and the opacity decreases. As Y is ionized further, up to V $-$ VI, the shell configuration resembles neutral $p$-shell elements ($Z = 35\,-\,34$) with an energy level distribution at a higher energy. The opacity peaks when Yttrium is sextuply ionized (VII) at $T\,\sim\,50000$ K, at which it has a similar structure to neutral Arsenic (As, $Z$ = 33), with a half-closed shell structure. Beyond this ionization (VIII $-$ XI), Y becomes similar to the neutral $p$ ($Z = 32 - 31$) and $d$ ($Z = 30 - 29$) shell elements. This leads to a decrease in the number of available energy levels, consequently reducing the opacity. 

The behavior of the $d$-shell element Cadmium (Cd, $Z=$ 48) is more straightforward (top right panel of \autoref{fig:op_ion_Pl}). As the temperature increases, it loses $d$-shell electrons. When the $d$-shell has a half-closed structure, the element reaches peak opacity. Then, the opacity decreases as more $d$-shell electrons are lost at higher temperature.

The $p$-shell element Iodine (I, $Z$ = 53) has a complicated variation in opacity with the temperature but can be explained in a similar way (bottom left panel of \autoref{fig:op_ion_Pl}). The opacity is high when I resembles elements with half-closed shells. Namely, the opacity peaks at $T\,\sim\,10^{4}$ K and at $T\,\sim\,10^{5}$ K, when I has a similar structure to the neutral $p$-shell element Antimony (Sb, $Z$ = 51) and $d$-shell element Technetium (Te, $Z$ = 43) respectively, being doubly (III) and tenth (XI) ionized. 

For the $s$-shell element Barium (Ba, $Z$ = 56, bottom right panel of \autoref{fig:op_ion_Pl}), the opacity reaches a peak at $T\,\sim\,4000$ K, when Ba is singly ionized (II) and has one neutral $s$ electron, similar to Caesium (Cs, $Z$ = 55). The opacity drops to a negligible value at $T\,\sim\,8000$ K, when doubly ionized Ba (III) resembles the energy level distribution of neutral $p$-shell element Xenon (Xe, $Z$ = 54), which has a closed $p$-shell. Such ions have most of their energy levels distributed at higher energies, and thus fewer transitions take place as the Boltzmann statistics predicts most electrons exist in the lower-lying energy levels at this temperature range. The opacity rises to a higher value at $T\,\sim\,30000$ K when the energy distribution is similar to the half-closed neutral $p$-shell element Sb ($Z$ = 51). As the ionization degree increases, the opacity decreases again when Ba resembles the configuration of neutral $d$-shell elements with lower complexity.

The variation of the Planck mean opacity with temperature for all the elements of interest can be understood in the same manner (\autoref{fig:op_Pl}).
With the increasing temperature and ionization, the effective shell structure of the ions change, the opacity varying accordingly. Highly ionized elements have the maximum bound-bound opacities when they have a half-closed shell structure.

For $d$-shell elements, opacity as a function of temperature peaks when it has half-closed $p$-shell or half-closed $d$-shell structures. The peak opacity is higher when the element has a half-closed $d$-shell structure ($\kappa_{\rm mean}\,\sim\, 1\, \rm cm^{2}\,g^{-1}$) rather than a half-closed $p$-shell structure ($\kappa_{\rm mean}\,\sim\, 0.1\, \rm cm^{2}\,g^{-1}$).
Most of the $p$-shell elements have $d$-shell electrons at high ionization, with the opacity peaking at $\kappa_{\rm mean}\,\sim\, 1\, \rm cm^{2}\,g^{-1}$.
This is the reason why at early times (higher temperature), $p$-shell elements have comparable opacity contributions to $d$-shell elements. 

The $s$-shell elements have comparatively lower opacity $\kappa_{\rm mean}\,\sim\,0.001\, - \,0.1\, \rm cm^{2}\,g^{-1}$.
This lower opacity can be explained by $s$-shell elements never resembling neutral half d-shell elements at higher ionization, although they can be similar to neutral half p-shell elements.

The behavior of opacity and temperature for different elements is summarized in \autoref{fig:ZvsTK}. At lower temperatures, the elements with the maximum number of low-lying energy levels have the maximum opacity. At high temperatures, ions lose their initial outermost electrons and effectively have different shell structures. Furthermore, the higher-level transitions become attainable at high temperatures since higher energy levels are populated. In this case, the maximum contribution to the opacity typically comes from ions which have half-closed shells, with the highest complexity measure.

\begin{figure}[t]
\begin{center}
\includegraphics[width=0.5\textwidth]{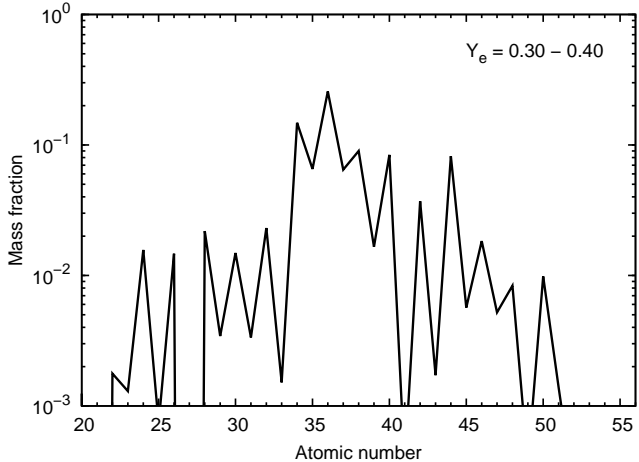}
\caption{The abundance pattern averaged for the electron fraction range of $Y_{\rm{e}}$ = 0.30 $-$ 0.40. 
}
\label{fig:abun_mix}
\end{center}
\end{figure}
\begin{figure*}[t]
\begin{center}
  \begin{tabular}{c}
    % 1(first figure)
    \begin{minipage}{0.5\hsize}
      \begin{center}
        \includegraphics[width=\linewidth]{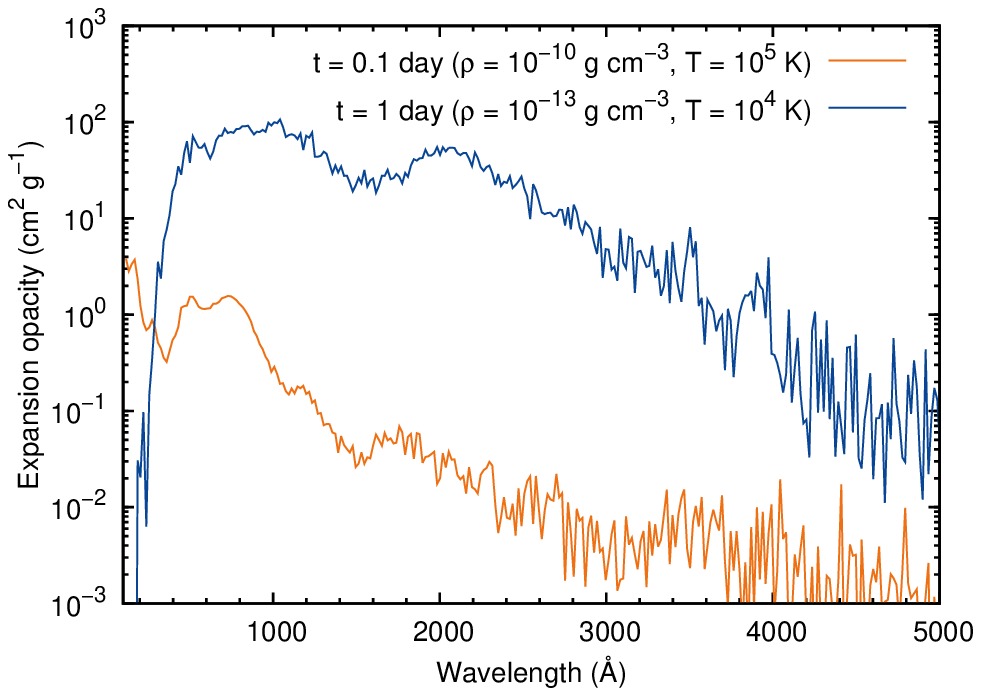} 
      \end{center}
    \end{minipage}

    \begin{minipage}{0.5\hsize}
      \begin{center}
          \includegraphics[width=\linewidth]{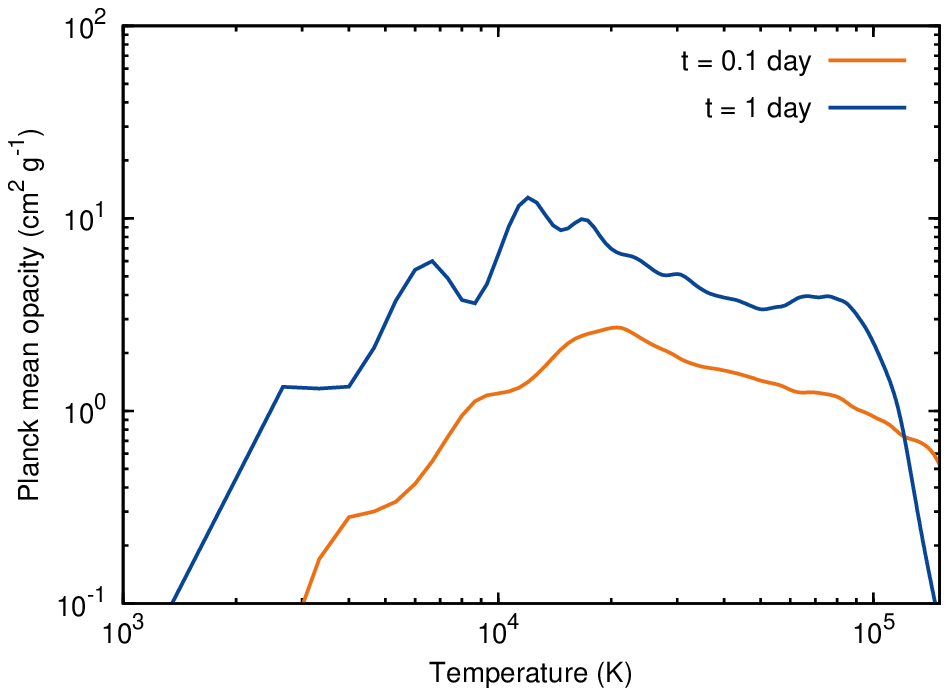}
      \end{center}
    \end{minipage}

  \end{tabular}
  \caption{\textbf{Left:} The expansion opacity as a function of wavelength for the mixture of elements. \textbf{Right:} The Planck mean opacity as a function of temperature for the mixture of elements. The orange and blue curves depict the condition at $t$ = 0.1 day and $t$ = 1 day for an electron fraction $Y_{\rm e}$ = 0.30 $-$ 0.40, and fixed densities $\rho\,=\,10^{-10}\,\rm g\,cm^{-3}$ and $\rho\, =\,10^{-13}\, \rm g\,cm^{-3}$, respectively.}
  \label{fig:opacity_mix}
  \end{center}
\end{figure*}

\subsubsection{Opacity of element mixture} \label{subsubsec:mix_bbop}

In this section, we consider the bound-bound opacities in the ejecta that consist of a mixture of different elements.
Depending on the electron fraction $Y_{\rm{e}}$, a different abundance pattern is realized in the ejecta. To estimate the bound-bound opacity for blue kilonova, we calculate the opacity for the mixture of elements in an ejecta, assuming $Y_{\rm{e}}\,=\,0.30\,-\,0.40$. 
We take the abundance pattern using the results from \citet{Wanajo2014}. We assume that the mass distribution in the each $Y_{\rm{e}}$ bin is flat.
At such high $Y_{\rm{e}}$, the second and third peak $r$-process elements are not synthesized. The elements with a significant abundance are $Z\,\sim 35 \,- \,45$ (\autoref{fig:abun_mix}).

The left panel of \autoref{fig:opacity_mix} shows the expansion opacity as a function of wavelength for the element mixture at $t\,=$ 0.1 and $t\,=$ 1 day. To model the typical conditions at these times, we set $\rho\,=\,10^{-10} \,\rm g\,cm^{-3}$, $T\,=\,10^{5}$ K for $t\,=$ 0.1 day; $\rho\,=\,10^{-13} \,\rm g\,cm^{-3}$, $T\,=\,10^{4}$ K for $t\,=$ 1 day.
At $t\,=$ 1 day, the expansion opacity peaks at $\kappa_{\rm exp}\,\sim\,10^{2}\,\rm{cm^{2}\,g^{-1}}$, whereas the peak expansion opacity at $t\,=$ 0.1 day only reaches $\kappa_{\rm exp}\,\sim\,1\,\rm{cm^{2}\,g^{-1}}$. 
The Planck mean opacity also shows an increase with time (right panel of \autoref{fig:opacity_mix}). The value of opacity is $\kappa_{\rm mean}\,\sim 0.5-1 \,\rm{cm^{2}\,g^{-1}}$ for the typical conditions at $t\,=$ 0.1 day; under the typical conditions at $t\,=$ 1 day, $\kappa_{\rm mean}\,\sim 5-10 \,\rm{cm^{2}\,g^{-1}}$.
These results can be understood using \autoref{eqn:kexp}.
Since the expansion opacity is inversely proportional to $\rho t$, the change in $\rho t$ from $t\,\sim$ 0.1 to 1 day increases the opacity by a factor of 100. Meanwhile, the Sobolev optical depth decreases with time, which reduces the contribution from the summation of $1-e^{-\tau_l}$. As a result, the opacity increases by a factor of about 10 as time increases from $t = $ 0.1 to 1 day.

The Planck mean opacity results for an element mixture
as a function of temperature
(right panel of \autoref{fig:opacity_mix}) can be understood
by individual element properties.
At relatively low temperatures ($T < 20000$ K), the opacity increases with temperature.
This is a property of $d$-shell elements that have the largest contribution to the opacity in this temperature range.
The opacity displays some modulation by reflecting
the behaviors of abundant individual elements.
At high temperatures ($T > 20000$ K), the opacity evolves more smoothly with temperature because the contributions from $p$- and $d$-shell elements with different peak positions in the Planck mean opacity are averaged out.

Hence, as evidenced by the results, the bound-bound opacity is orders of magnitude greater than the electron scattering, bound-free and free-free opacities. At $t\,=$ 0.1 day, the Planck mean of bound-bound opacity can reach up to a value of $\kappa_{\rm mean}\,\sim 2\,\rm{cm^{2}\,g^{-1}}$, whereas other contributions to the total opacity, $\kappa^{\rm{es}}\,=\,(3\,-10)\,\times10^{-2} \,\rm{cm^2\,g^{-1}}$ and $\kappa^{\rm{ff}}_{i,j}=(2\,-\,3)\,\times10^{-4}\,\rm{cm^2\,g^{-1}}$, are negligible at a wavelength $\lambda$ = 1000 {\AA} for $Z = 20 - 56$ (\autoref{sec:opacity}). The bound-free opacity is not significant at this time since the fraction of photons with energy beyond the photo-ionization threshold is small (\autoref{sec:opacity}). Therefore, we conclude that bound-bound opacity is the most significant component of the total opacity at an early time ($t \sim 0.1$ days).

\section{Radiative transfer simulations}  \label{sec:radtransfer}
Using the new atomic data and opacities, we calculate the light curve of blue kilonova using a time-dependent and wavelength-dependent radiative transfer code \citep{tanaka2013, tanaka2014, tanaka2017, Kawaguchi2018}. With a given density structure and $Y_{\rm{e}}$ distribution, the code calculates the light curves and spectra. The radioactive heating rate of $r$-process nuclei is calculated according to $Y_{\rm{e}}$, using the results from \citet{Wanajo2014}. The photon transfer is calculated by a Monte Carlo method. The time-dependent thermalization factor is adopted from \citet{Barnes2016}. The new opacity data enables us to calculate the radiation transfer starting around $\sim\, 1$ hour after the merger. We consider the transitions in a
wavelength range 100 $\rm \AA$ - 35000 \AA. The simulation is performed from 0.03 to 300 days to calculate the light curves.
We describe our model in \autoref{subsec:model} and discuss the evolution of opacity in the ejecta in \autoref{subsec:op_ev}. Our results for the bolometric luminosity calculation using this opacity is presented in \autoref{subsec:l_bol}.
\subsection{Model}\label{subsec:model}
We use a simple ejecta model \citep{Metzger2010} which considers a spherical ejecta expanding homologously. As our fiducial case, we use the power-law density structure $\rho \propto r^{-3}$ from a velocity $v\,=$ 0.05c to 0.2c,
a total ejecta mass of $M_{\rm ej}\, =\,0.05M_{\odot}$, and an electron fraction range of $Y_{\rm{e}}\,=$ 0.30 $-$ 0.40. Similarly to \autoref{subsubsec:bb_op}, we assume a flat distribution of mass for each value in the $Y_{\rm{e}}$ range, subsequently using the results from \citet{Wanajo2014} to calculate the abundance pattern. Throughout the ejecta, the same $Y_{\rm{e}}$ distribution, and hence homogeneous elemental abundance pattern, are assumed. The velocity scale and the range of $Y_{\rm{e}}$ in our fiducial model are typical for disk wind ejecta, particularly in the case of a relatively long-lived hypermassive NS \citep{Perego2014,Metzger2014,Lippuner2017,Siegel2017,Fujibayashi2018,Fernandez2019}. In such conditions, the main nucleosynthesis products are light $r$-process elements (\autoref{fig:abun_mix}).

In reality, the disk wind ejecta are enveloped inside a faster moving dynamical ejecta \citep{Hotokezaka2013a}. To study the effect of this dynamical ejecta, we further include models with a continuous thin outer layer at $v\,>\,0.2c$ with a fixed mass of $M_{\rm{out}}\,=\,0.005M_{\odot}$. The layer has a steeper density structure $\rho \propto r^{n}$ where $n = -6, -8$, and $-10$. According to the slope, the maximum outer velocity changes as $v\,\sim$ 0.24c, 0.25c, and 0.33c, for $n\,= -6, -8$, and $-10$, respectively. 
We assume the same $Y_{\rm{e}}$ range for these outer ejecta components. These modelling conditions may be applicable for a 
shock-heated polar dynamical ejecta, where $Y_{\rm{e}}$ can rise by $e^{+}$ capture and $\nu_{e}$ absorption  \citep{Goriely2015, Sekiguchi2015, Sekiguchi2016,Martin2018, Radice2018}. Thus, even with relatively high $Y_{\rm{e}}$, our model can provide a sound approximation for the emission viewed from the polar direction. We do not include lanthanide-rich ejecta as the main focus of this work is to present the light curves of blue kilonovae. 

\subsection{Evolution of opacity}\label{subsec:op_ev}
\begin{figure}[t]
       \includegraphics[width=\linewidth]{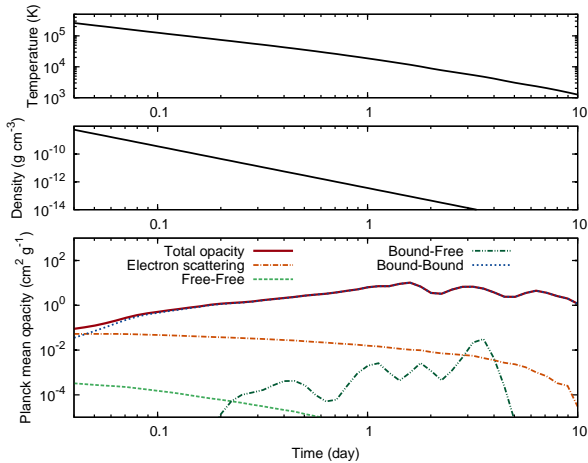}

  \caption{\textbf{Top:} The temperature evolution for the fiducial model with $M_{\rm ej} = 0.05M_{\odot}$ and $Y_{\rm e}$ = 0.30 $-$ 0.40, at a fixed ejecta point $v$ = 0.1c. \textbf{Middle:} The density evolution for the fiducial model at $v$ = 0.1c. \textbf{Bottom:} The Planck mean opacity variation with time at $v$ = 0.1c. The red line describes the total opacity. The blue line describes the bound-bound opacity, the component which contributes the most to the total opacity except for around $t \sim 1$ hour. The orange, light green, and the dark green curves are the electron scattering, free-free, and bound-free opacity components, respectively. 
  }
  \label{fig:opacity_time}
%   \end{center}
\end{figure}

As the ejecta expands, the temperature and the density of the ejecta decrease. The opacity also evolves with time accordingly. Therefore, it is useful to study the time evolution of the opacities at a fixed position in the ejecta. \autoref{fig:opacity_time} shows the temperature, density and opacity evolution at the ejecta point $v\, =\, 0.1c$ for our fiducial model.
The dominant component is the bound-bound opacity, followed by the electron-scattering, bound-free, and free-free opacity. The total opacity varies from 0.1 $-$ 10 $\rm cm^{2}\,g^{-1}$ with time.

The contribution of electron scattering to the total opacity is higher at earlier times, reaching a majority contribution, $>\,50 \,\%$, at $t\,\sim$ 1 hour (\autoref{fig:opacity_time}).
This high electron scattering contribution occurs at an early time as high temperatures ($T\,>\,10^{5}$ K) cause a high degree of ionization, which raises electron density. The electron scattering opacity decreases with time as the ejecta temperature decreases. 
Around $t\,\sim 6$ days, the electron scattering contribution drops steeply because most of the elements recombine to neutral atoms.

The free-free component remains small throughout the evolution of the total opacity. At $t\sim\,0.1$ day, 
the opacity has a value of $\kappa^{\rm{ff}}_{\rm{mean}}\sim\,10^{-4}\,\rm cm^{2}\,g^{-1}$
; this value falls faster than the electron scattering opacity component as time increases.

\begin{figure}[t]\label{fig:fidu_lbol}
\includegraphics[width=\linewidth]{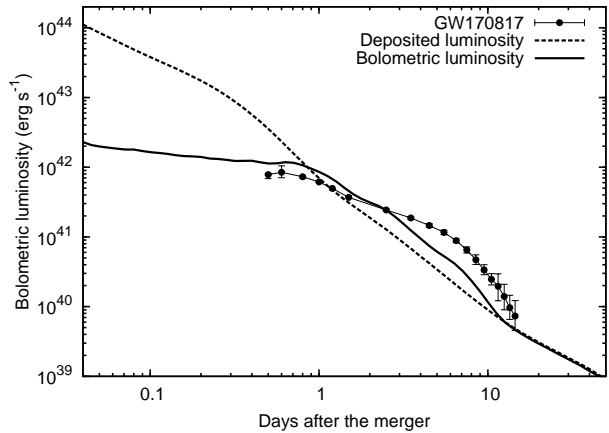}
\caption{The bolometric light curve for the fiducial model with $M_{\rm ej} = 0.05M_{\odot}$ and $Y_{\rm e}$ = 0.30 $-$ 0.40 and $\rho\, \propto\,r^{-3}$. The black dots represent the bolometric light curve of GW170817 \citep{Waxman2018}. \label{fig:Lbol_fidu}}
\end{figure}

(\autoref{fig:opacity_time}).
From \autoref{eqn:eqn_ff} and \autoref{eqn:ion_fr}, we can see that the free-free opacity varies as $\kappa^{\rm{ff}}_{i,j}(\lambda)\,\propto\, \rho\,T^{-\frac{1}{2}} (1 - e^{-\frac{hc}{\lambda kT}})$.
Since the density decreases faster than temperature as time increases, the free-free opacity decreases with time.

The bound-free opacity varies with time but never becomes large enough to significantly contribute to the total opacity.
As discussed in \autoref{subsec:bf}, although the photo-ionization cross section itself is high, the fraction of high energy photons is small, and thus, the Planck mean opacity is moderate.
The bound-free opacity component shows an increasing trend, reaching its peak value of $\kappa_{\rm mean}^{\rm bf}\,\sim\, 0.04\,\rm cm^{2}\,g^{-1}$ a few days after the merger.
This is as a result of the ionization degree decreasing with time, hence more photons are present beyond the potential energy of ions.
It should be noted that the value of the bound-free opacity before $t$ = 0.2 day is not correctly followed in the radiative transfer code, since the wavelength range beyond the ionization threshold is not covered by our wavelength grid (down to 100 \AA).

The bound-bound opacity component evolves from $\kappa_{\rm mean}^{\rm bb}\,\sim\, 0.5 $ to $\, 5 \,\rm cm^{2}\,g^{-1}$ from $t=$ 0.1 to 1 day. Excluding the time around 1 hour, this component alone is representative of the total opacity. It is to be noted that most of the previous works have considered a fixed opacity value of 1 $\rm cm^{2}\,g^{-1}$ or less \citep{mansi2017,Villar2017, Piro2018, Gottlieb2020} to calculate blue kilonovae at $t \,<\,1$ day. This assumption is not valid precisely, as the change in the opacity with time is quite large for even high $Y_{\rm e}$ ejecta.

\subsection{Bolometric light curves}\label{subsec:l_bol}

The bolometric luminosity for the fiducial model is shown in \autoref{fig:Lbol_fidu}. The luminosity deposited the ejecta (or thermalized radioactive luminosity) is shown by the dashed line for comparison.
At $t\,<$ 1 day, the observable bolometric luminosity is an order of magnitude lower than the deposition luminosity because the ejecta are optically thick, hence photons cannot escape from the ejecta.
At $t\,>$ 1 day, the previously stored radiation energy from $t\,<$ 1 day starts to be released and the bolometric luminosity supersedes the deposition luminosity.
Finally, the bolometric luminosity follows the thermalized radioactive emission at $t\,>$ 10 days.

The bolometric light curve of GW170817 \citep{Waxman2018} is shown for comparison. Our fiducial model with $M_{\rm ej} = 0.05M_{\odot}$ gives a reasonable agreement with the observed data at early times. The required ejecta mass is consistent with the findings of previous works \citep{mansi2017,Waxman2018,Hotokezaka2020}.

\begin{figure}[t]
    % \centering
\includegraphics[width=\linewidth]{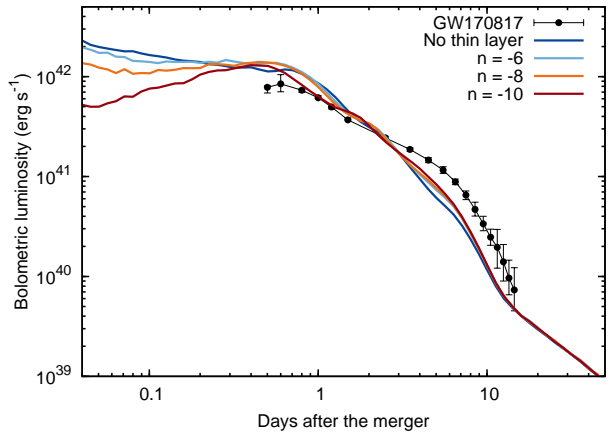}
\caption{The bolometric light curve for models containing a thin layer, with slopes $n$ = $-6$, $-8$, $-10$. The light curve becomes fainter at early times with the inclusion of the steeper thin outer layer. The bolometric light curves of GW170817 \citep{Waxman2018} are shown for reference.  \label{fig:Lbol_slope}}
\end{figure}

The presence of a thin outer layer affects the light curves at an early time (\autoref{fig:Lbol_slope}).
The steeper slope of the outer ejecta makes the luminosity fainter at $t \leq$ 1 day.
In the early time, the ejecta are optically thick and the emission from the outermost layer determines the light curve. Adding a thin outer layer to the ejecta changes the mass located outside of the diffusion sphere in the early time. Our fiducial model has a higher density at the diffusion sphere, producing a high luminosity in the early time (\autoref{fig:Lbol_slope}). For the models with thin layers, the density at the diffusion sphere becomes lower. Since the model with a steeper slope has a lower density of the optically thin layer for a fixed mass of the outer ejecta, the model displays a fainter luminosity.
After around $t > 1$ day, the thin layer has almost no effect on the light curve
because thin ejecta are already optically thin and so do not contribute to the luminosity anymore.

\begin{figure*}[t]
\begin{tabular}{c}
\begin{minipage}{0.5\hsize}
\begin{center}
\includegraphics[width=\linewidth]{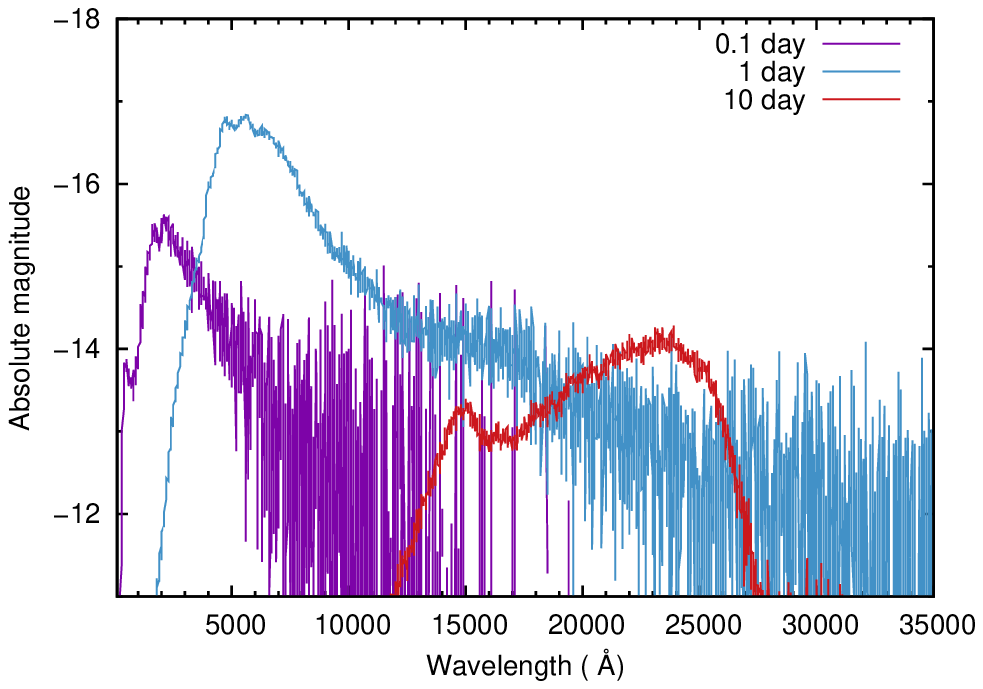} 
\end{center}
\end{minipage} 
\begin{minipage}{0.5\hsize}
\begin{center}

\includegraphics[width=\linewidth]{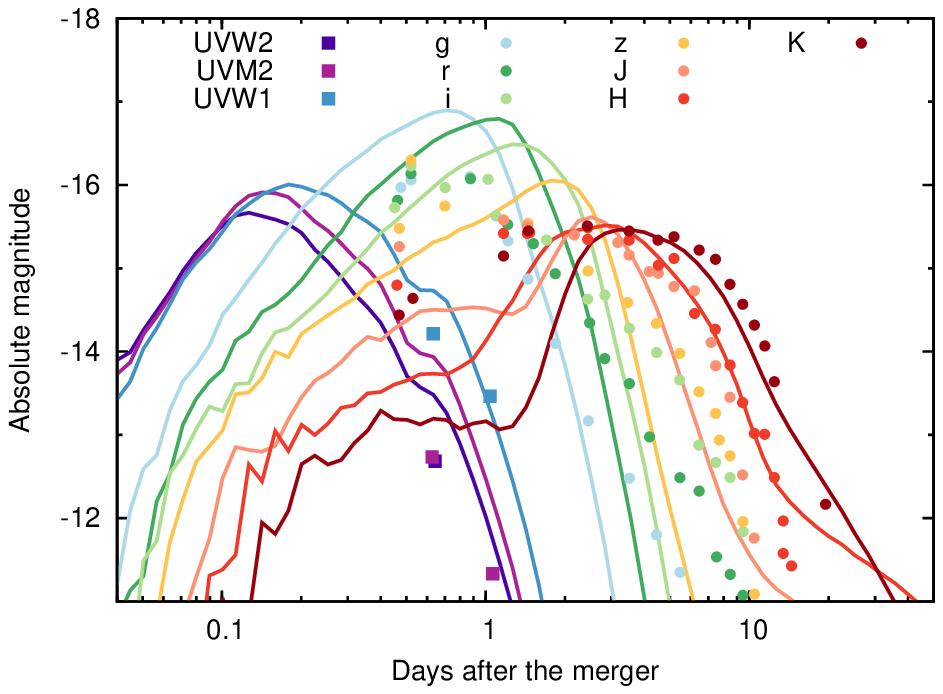} 
\end{center}
\end{minipage}    
    
\end{tabular}{} 
\caption{The spectra and light curves for the fiducial model, where $M_{\rm ej} = 0.05M_{\odot}$ and $Y_{\rm e}$ = 0.30 $-$ 0.40 for an assumed density of $\rho \propto r^{-3}$. The light curves are shown for $t\,=$ 0.04 $-$ 50 days. \textbf{Left panel:} The spectra 
($f_{\nu}$ flux shown as absolute AB magnitude) at different times after the merger: with the purple line at 0.1 day, the blue line at 1 day, and the red line at 10 days. The spectra evolve from UV towards the NIR. \textbf{Right panel:} Multi-color light curves of the model compared with the data of GW170817 (UV data from \citet{Evans2017, Drout2017} and the other data compiled by \citet{Villar2017}).}
\label{fig:multi_lc}
\end{figure*}

\section{Discussions}\label{sec:discussion}
\subsection{Applications to GW170817}\label{subsec:multi_gw170817}

The spectra at $t=\,0.1, \,1,$ and 10 days after the merger, and the multi-color light curves, both plotted in absolute magnitude, are shown for the fiducial model (left and right panel in \autoref{fig:multi_lc}). The spectral shape shows a strong time evolution from shorter wavelengths ($\lambda\,\leq$ 10000 \AA $\,$) towards infrared wavelengths ($\lambda \geq 10000$ \AA) at later times.
This trend is also displayed in the multi-color light curves.
The peak times of the light curves gradually move from shorter to longer wavelengths: the UV light curves peak at $t\,\sim$ 4.3 hours,
blue optical light curves peak at $t\,\sim$ 16.8 hours,
and red optical and NIR light curves peak at $t > $ 1 day.
The early UV emission declines very quickly and becomes fainter than an absolute magnitude of $-10$ in $t\,\le$ 2 days.
A similar pattern for blue optical emission occurs over a somewhat longer timescale. The NIR brightness remains bright from 1 day to a week. 

We compare our model with the multi-color light curves of GW170817 \citep{Villar2017}. The data are corrected for Galactic extinction with $E(B-V)=0.1$. This comparison provides insight on the emission mechanism of GW170817-like events.
Similar to the bolometric luminosity (\autoref{fig:Lbol_fidu}), our fiducial model shows reasonable agreement with the data. 
Hence, our model shows that a one-component, purely-radioactive, high $Y_{\rm e}$ ejecta can explain early-time bright UV and blue emission. Previous studies that assumed a constant opacity also show good agreement for early UV and blue optical data \citet{Cowperthwaite2017,Drout2017,kasen2017,Villar2017}.
However, our calculations directly calculate atomic opacities, and thus, the opacity is not a free parameter in our model.

The UV magnitudes become fainter and decline faster upon the inclusion of a thin layer outside the fiducial model ejecta (\autoref{fig:mag_thl}); also pointed out by \citet{kasen2017}.
The UV light curves are shown for the fiducial model and the case where $n = - 10$. The UVW1 magnitude of the fiducial model without a thin layer peaks at an absolute magnitude of $- 16$ mag at $t\,\sim\,4.3$ hours, whereas that of the model incorporating a thin layer with $n = -10$ is fainter at $t\,> \,0.1$ day, reaching a peak of $-15.7$ mag at $t\,\sim\,2.4$ hours.

It should be noted that our models assume that the outer ejecta are lanthanide-free, with $Y_{\rm e} = 0.30-0.40$.
If the outer ejecta have a lower $Y_{\rm e}$, as expected for dynamical ejecta in the equatorial plane, the UV brightness can be suppressed further. 
Hence, the purely radioactive kilonova models may not be able to explain the observed early light curve, depending on the structure and composition of the outer ejecta.
In this case, a heating source other than radioactive decays of $r$-process nuclei may be necessary, for example, heating by shock or cocoon \citep{mansi2017,Piro2018}, $\beta$ decay luminosity from free neutrons \citep{Metzger2015,Gottlieb2020},
or some other central power source \citep{Metzger2008, Yu2013, Metzger2014,  Matsumoto2018,Metzger2018,Li2018,Wollaeger2019}.
Although it is difficult to draw firm conclusions due to a lack of atomic data on highly-ionized lanthanides, 
our new atomic opacities provide the foundation for a discussion on the detailed properties of blue kilonova models at an early phase.

\subsection{Future prospects}\label{subsec:future_pros}

\begin{figure}[t]
\centering
 \includegraphics[width=\linewidth]{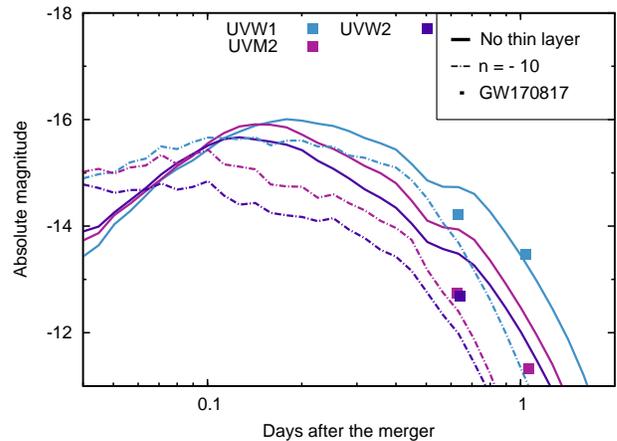}
    
  \caption{The comparison of UV magnitudes between the fiducial model ejecta with density structure $\rho \propto r^{-3}$, $M_{\rm ej} = 0.05M_{\odot}$, and $Y_{\rm e}$ = 0.30 $-$ 0.40, and the model with a thin outer layer with a slope $n$ = $-10$. The magnitude becomes fainter with the inclusion of a steeper outer thin layer. The data of GW170817 \citep{Evans2017, Drout2017} are shown in squares for comparison.}
  \label{fig:mag_thl}

\end{figure}

\begin{figure*}[t]
\begin{center}
  \begin{tabular}{c}
   
     \begin{minipage}{0.5\hsize}
      \begin{center}
        \includegraphics[width=\textwidth]{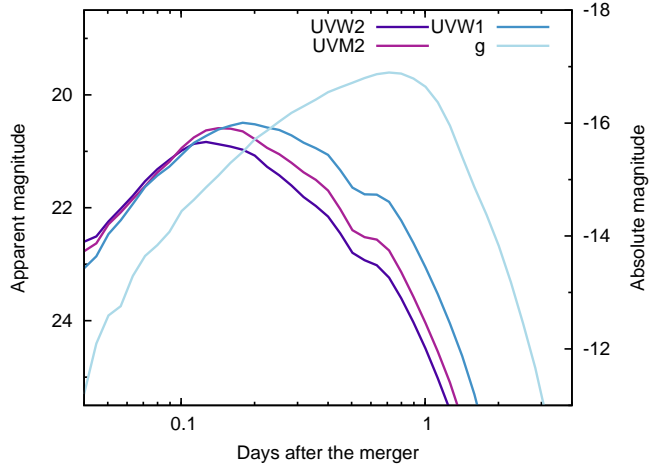} 
      \end{center}
    \end{minipage}
    
  \end{tabular}
  \caption{
  The apparent UV and optical $g$-band light curves
  for the fiducial model at 200 Mpc distance.}
  \label{fig:mag_scaled}
  \end{center}
\end{figure*}

Finally, we discuss the prospect of observing an early kilonova emission.
Our simulations give the first synthetic light curves of kilonova at a timescale of hours after the merger, based on the detailed atomic opacities.
As discussed in \autoref{subsec:multi_gw170817}, the early UV emission is sensitive to the structure of the ejecta.
Furthermore, contributions from other heating sources may play important roles determining the early luminosity.
Therefore, the early-time observations provide clues to distinguish these models.

Although the first UVOIR data of GW170817 were obtained at 11 hours,
the early data suggest that UV and blue optical emissions peak before this time.
In our fiducial model, UVW1 and optical $g$-band magnitudes peak at $t\,\sim$ 4.3 hours and 16.8 hours respectively (\autoref{fig:multi_lc}).
When the ejecta is enveloped by a thin outer layer, with a density slope $n = -10$, the UVW1 magnitude peaks earlier, reaching a value of $-15.7$ mag at $t\,\sim$ 2.4 hours (\autoref{fig:mag_thl}). \autoref{fig:mag_scaled} shows the expected observed magnitudes in UV and optical $g$-band at 200 Mpc.
The UVW1 and $g$-band magnitudes at 200 Mpc reach apparent magnitudes of $\sim$ 20.5 and 19.8 mag, respectively at the peak time.

The early UV signals are bright enough to be detected by existing facilities like Swift \citep{Roming2005}, which has a limiting magnitude of 22 mag for an exposure time of 1000 s \citep{Brown2014}, if a counterpart is discovered early enough to start UV observations promptly.
A more promising detection method is via wide-field survey observations in the UV wavelengths, as UV emissions peak earlier than optical emissions.
Upcoming wide field surveys, such as those carried out by the Ultraviolet Transient Astronomy Satellite (ULTRASAT, \citealt{Sagiv2014}), which can detect down to the AB magnitude of about 22.4 mag in 900 s, are able to detect the expected signal from our fiducial model even at 500 Mpc distance.

\section{Conclusions}\label{sec:conclusion}

In this paper, we have calculated the atomic structures and opacities for the elements with atomic number of $Z\,= 20 - 56$, which are necessary to understand the properties of early blue kilonova from NS mergers.
We cover ionizations up to the tenth degree (XI) at a typical ejecta temperature at $t=0.1$ days ($T\,=\,10^{5}$ K).
We find that the bound-bound opacities are the dominant source of the opacity at early times ($t\,<\,1$ day).
Among different elements and ionization states, ions with half-closed electron shells provide the highest contributions to the bound-bound opacity.
The Planck mean opacity of the lanthanide-free ejecta at early times is about one order of magnitude lower than the opacity at late times: $\kappa\,\sim 0.5-1 \,\rm{cm^{2}\,g^{-1}}$ at  $t\,\sim$ 0.1 day, compared to $\kappa\,\sim 5-10 \,\rm{cm^{2}\,g^{-1}}$ at $t\,\sim$ 1 day. 

Using this opacity, we have performed multi-wavelength radiative transfer simulations and calculated the bolometric and multi-color light curves of blue kilonova.
Our fiducial model, with an ejecta mass of $M_{\rm ej} = 0.05M_{\odot}$,
reaches the bolometric luminosity of $\sim\,2\,\times\,10^{42}\,\rm erg\,s^{-1}$ by $t = 0.1$ days.
The UV and blue optical band magnitudes reach their peak absolute magnitudes of $-16$ mag and $-17$ mag at $t\,\sim$ 4.3 and 16.8 hours, respectively. The behaviors of early light curves are affected by the outer structure of the ejecta. The presence of a thin outer layer greatly suppresses the luminosity at $t\,\leq$ 1 day, agreeing with the results of \citet{kasen2017}.
The comparison of our fiducial model with the bolometric and multi-color light curves of GW170817 in the early phase shows reasonable agreement. Our result suggests that the early data of GW170817 can be explained with a purely radioactive kilonova model with a high $Y_{\rm e}$ ejecta.

However, there are some limitations of our models. Firstly, we have considered only high $Y_{\rm e}$, lanthanide-free outer ejecta. As the low $Y_{\rm e}$, lanthanide-rich ejecta can have quite different properties of opacities, which in turn affect the luminosity, we cannot yet firmly exclude the possibility of a heating source other than radioactive heating at early times.
Moreover, we did not take into account the multi-dimensional, multi-component ejecta, as considered by e.g., \citet{Villar2017}, \citet{ Perego2017}, and \citet{Kawaguchi2018}.
These assumptions prevent us from precisely predicting the emission viewed from the equatorial region, where the presence of the lanthanide-rich, dynamical ejecta is expected.

Despite these limitations, our model will be helpful in quantitatively comparing the models with future early-time observations for NS merger events.
Our model predicts bright early-time UV emission that is detectable with 
the Swift satellite at a distance of 200 Mpc, if observations are started promptly.
Furthermore, wide-field UV surveys with upcoming satellites such as ULTRASAT can detect such emissions, even at distances as large as 500 Mpc.
Such early UV observations will provide rich information about the structure of the outermost ejecta, providing a unique way to test the presence of other heating sources such as a cocoon, decays of free neutrons, or a long-lived central engine.

\acknowledgments

We want to express our sincere gratitude to the anonymous referee for providing us with constructive comments. Numerical simulations presented in this paper 
were carried out with Cray XC50 at the Center for Computational Astrophysics,
National Astronomical Observatory of Japan.
This research was supported by the JSPS Bilateral Joint Research Project
and the Grant-in-Aid for Scientific Research from
JSPS (16H02183, 19H00694, 20H00158) and MEXT (17H06363).

\vspace{5mm}

\bibliography{bibfolder.bib}{}

\begin{thebibliography}{}
\expandafter\ifx\csname natexlab\endcsname\relax\def\natexlab#1{#1}\fi
\providecommand{\url}[1]{\href{#1}{#1}}
\providecommand{\dodoi}[1]{doi:~\href{http://doi.org/#1}{\nolinkurl{#1}}}
\providecommand{\doeprint}[1]{\href{http://ascl.net/#1}{\nolinkurl{http://ascl.net/#1}}}
\providecommand{\doarXiv}[1]{\href{https://arxiv.org/abs/#1}{\nolinkurl{https://arxiv.org/abs/#1}}}

\bibitem[{{Abbott} {et~al.}(2017){Abbott}, {Abbott}, {Abbott}, {Acernese},
  {Ackley}, {Adams}, {Adams}, {Addesso}, {Adhikari}, {Adya}, {Affeldt},
  {Afrough}, {Agarwal}, {Agathos}, {Agatsuma}, {Aggarwal}, {Aguiar}, {Aiello},
  {Ain}, {Ajith}, {LIGO Scientific Collaboration}, \& {Virgo
  Collaboration}}]{Abbott2017a}
{Abbott}, B.~P., {Abbott}, R., {Abbott}, T.~D., {et~al.} 2017, \prl, 119,
  161101, \dodoi{10.1103/PhysRevLett.119.161101}

\bibitem[{{Arcavi}(2018)}]{arcavi2018}
{Arcavi}, I. 2018, ApJ letters, 855, L23.
\newblock \doarXiv{1802.02164}

\bibitem[{{Bar-Shalom} {et~al.}(2001){Bar-Shalom}, {Klapisch}, \&
  {Oreg}}]{bar-shalom01}
{Bar-Shalom}, A., {Klapisch}, M., \& {Oreg}, J. 2001, \jqsrt, 71, 169,
  \dodoi{10.1016/S0022-4073(01)00066-8}

\bibitem[{{Barnes} {et~al.}(2016){Barnes}, {Kasen}, {Wu}, \&
  {Mart{\'\i}nez-Pinedo}}]{Barnes2016}
{Barnes}, J., {Kasen}, D., {Wu}, M.-R., \& {Mart{\'\i}nez-Pinedo}, G. 2016,
  \apj, 829, 110, \dodoi{10.3847/0004-637X/829/2/110}

\bibitem[{{Brown} {et~al.}(2014){Brown}, {Breeveld}, {Holland}, {Kuin}, \&
  {Pritchard}}]{Brown2014}
{Brown}, P.~J., {Breeveld}, A.~A., {Holland}, S., {Kuin}, P., \& {Pritchard},
  T. 2014, \apss, 354, 89, \dodoi{10.1007/s10509-014-2059-8}

\bibitem[{{Connaughton} {et~al.}(2017){Connaughton}, {Goldstein}, \& {Fermi GBM
  - LIGO Group}}]{Connaughton2017}
{Connaughton}, V., {Goldstein}, A., \& {Fermi GBM - LIGO Group}. 2017, in
  American Astronomical Society Meeting Abstracts, Vol. 229, American
  Astronomical Society Meeting Abstracts \#229, 406.08

\bibitem[{{Coulter} {et~al.}(2017){Coulter}, {Foley}, {Kilpatrick}, {Drout},
  {Piro}, {Shappee}, {Siebert}, {Simon}, {Ulloa}, {Kasen}, {Madore},
  {Murguia-Berthier}, {Pan}, {Prochaska}, {Ramirez-Ruiz}, {Rest}, \&
  {Rojas-Bravo}}]{Coulter2017}
{Coulter}, D.~A., {Foley}, R.~J., {Kilpatrick}, C.~D., {et~al.} 2017, Science,
  358, 1556, \dodoi{10.1126/science.aap9811}

\bibitem[{{Cowperthwaite} {et~al.}(2017){Cowperthwaite}, {Berger}, {Villar},
  {Metzger}, {Nicholl}, {Chornock}, {Blanchard}, {Fong}, {Margutti},
  {Soares-Santos}, {Alexander}, {Allam}, {Annis}, {Brout}, {Brown}, {Butler},
  {Chen}, {Diehl}, {Doctor}, {Drout}, {Eftekhari}, {Farr}, {Finley}, {Foley},
  {Frieman}, {Fryer}, {Garc{\'\i}a-Bellido}, {Gill}, {Guillochon}, {Herner},
  {Holz}, {Kasen}, {Kessler}, {Marriner}, {Matheson}, {Neilsen}, {Quataert},
  {Palmese}, {Rest}, {Sako}, {Scolnic}, {Smith}, {Tucker}, {Williams},
  {Balbinot}, {Carlin}, {Cook}, {Durret}, {Li}, {Lopes}, {Louren{\c{c}}o},
  {Marshall}, {Medina}, {Muir}, {Mu{\~n}oz}, {Sauseda}, {Schlegel}, {Secco},
  {Vivas}, {Wester}, {Zenteno}, {Zhang}, {Abbott}, {Banerji}, {Bechtol},
  {Benoit-L{\'e}vy}, {Bertin}, {Buckley-Geer}, {Burke}, {Capozzi}, {Carnero
  Rosell}, {Carrasco Kind}, {Castander}, {Crocce}, {Cunha}, {D'Andrea}, {da
  Costa}, {Davis}, {DePoy}, {Desai}, {Dietrich}, {Drlica-Wagner}, {Eifler},
  {Evrard}, {Fernand ez}, {Flaugher}, {Fosalba}, {Gaztanaga}, {Gerdes},
  {Giannantonio}, {Goldstein}, {Gruen}, {Gruendl}, {Gutierrez}, {Honscheid},
  {Jain}, {James}, {Jeltema}, {Johnson}, {Johnson}, {Kent}, {Krause}, {Kron},
  {Kuehn}, {Nuropatkin}, {Lahav}, {Lima}, {Lin}, {Maia}, {March}, {Martini},
  {McMahon}, {Menanteau}, {Miller}, {Miquel}, {Mohr}, {Neilsen}, {Nichol},
  {Ogando}, {Plazas}, {Roe}, {Romer}, {Roodman}, {Rykoff}, {Sanchez},
  {Scarpine}, {Schindler}, {Schubnell}, {Sevilla-Noarbe}, {Smith}, {Smith},
  {Sobreira}, {Suchyta}, {Swanson}, {Tarle}, {Thomas}, {Thomas}, {Troxel},
  {Vikram}, {Walker}, {Wechsler}, {Weller}, {Yanny}, \&
  {Zuntz}}]{Cowperthwaite2017}
{Cowperthwaite}, P.~S., {Berger}, E., {Villar}, V.~A., {et~al.} 2017, \apjl,
  848, L17, \dodoi{10.3847/2041-8213/aa8fc7}

\bibitem[{{Drout} {et~al.}(2017){Drout}, {Piro}, {Shappee}, {Kilpatrick},
  {Simon}, {Contreras}, {Coulter}, {Foley}, {Siebert}, {Morrell}, {Boutsia},
  {Di Mille}, {Holoien}, {Kasen}, {Kollmeier}, {Madore}, {Monson},
  {Murguia-Berthier}, {Pan}, {Prochaska}, {Ramirez-Ruiz}, {Rest}, {Adams},
  {Alatalo}, {Ba{\~n}ados}, {Baughman}, {Beers}, {Bernstein}, {Bitsakis},
  {Campillay}, {Hansen}, {Higgs}, {Ji}, {Maravelias}, {Marshall}, {Moni Bidin},
  {Prieto}, {Rasmussen}, {Rojas-Bravo}, {Strom}, {Ulloa},
  {Vargas-Gonz{\'a}lez}, {Wan}, \& {Whitten}}]{Drout2017}
{Drout}, M.~R., {Piro}, A.~L., {Shappee}, B.~J., {et~al.} 2017, Science, 358,
  1570, \dodoi{10.1126/science.aaq0049}

\bibitem[{{Eastman} \& {Pinto}(1993)}]{Eastman1993}
{Eastman}, R.~G., \& {Pinto}, P.~A. 1993, \apj, 412, 731,
  \dodoi{10.1086/172957}

\bibitem[{{Eichler} {et~al.}(1989){Eichler}, {Livio}, {Piran}, \&
  {Schramm}}]{Eichler1989}
{Eichler}, D., {Livio}, M., {Piran}, T., \& {Schramm}, D.~N. 1989, \nat, 340,
  126, \dodoi{10.1038/340126a0}

\bibitem[{{Evans} {et~al.}(2017){Evans}, {Cenko}, {Kennea}, {Emery}, {Kuin},
  {Korobkin}, {Wollaeger}, {Fryer}, {Madsen}, {Harrison}, {Xu}, {Nakar},
  {Hotokezaka}, {Lien}, {Campana}, {Oates}, {Troja}, {Breeveld}, {Marshall},
  {Barthelmy}, {Beardmore}, {Burrows}, {Cusumano}, {D'A{\`\i}}, {D'Avanzo},
  {D'Elia}, {de Pasquale}, {Even}, {Fontes}, {Forster}, {Garcia}, {Giommi},
  {Grefenstette}, {Gronwall}, {Hartmann}, {Heida}, {Hungerford}, {Kasliwal},
  {Krimm}, {Levan}, {Malesani}, {Melandri}, {Miyasaka}, {Nousek}, {O'Brien},
  {Osborne}, {Pagani}, {Page}, {Palmer}, {Perri}, {Pike}, {Racusin}, {Rosswog},
  {Siegel}, {Sakamoto}, {Sbarufatti}, {Tagliaferri}, {Tanvir}, \&
  {Tohuvavohu}}]{Evans2017}
{Evans}, P.~A., {Cenko}, S.~B., {Kennea}, J.~A., {et~al.} 2017, Science, 358,
  1565, \dodoi{10.1126/science.aap9580}

\bibitem[{{Fernandez} \& {Metzger}(2014)}]{Fernandez2014}
{Fernandez}, R., \& {Metzger}, B. 2014, in AAS/High Energy Astrophysics
  Division \#14, AAS/High Energy Astrophysics Division, 304.07

\bibitem[{{Fern{\'a}ndez} {et~al.}(2019){Fern{\'a}ndez}, {Tchekhovskoy},
  {Quataert}, {Foucart}, \& {Kasen}}]{Fernandez2019}
{Fern{\'a}ndez}, R., {Tchekhovskoy}, A., {Quataert}, E., {Foucart}, F., \&
  {Kasen}, D. 2019, \mnras, 482, 3373, \dodoi{10.1093/mnras/sty2932}

\bibitem[{{Fontes} {et~al.}(2020){Fontes}, {Fryer}, {Hungerford}, {Wollaeger},
  \& {Korobkin}}]{Fontes2020}
{Fontes}, C.~J., {Fryer}, C.~L., {Hungerford}, A.~L., {Wollaeger}, R.~T., \&
  {Korobkin}, O. 2020, \mnras, 493, 4143, \dodoi{10.1093/mnras/staa485}

\bibitem[{{Fontes} {et~al.}(2017){Fontes}, {Fryer}, {Hungerford}, {Wollaeger},
  {Rosswog}, \& {Berger}}]{Fontes2017}
{Fontes}, C.~J., {Fryer}, C.~L., {Hungerford}, A.~L., {et~al.} 2017, arXiv
  e-prints, arXiv:1702.02990.
\newblock \doarXiv{1702.02990}

\bibitem[{{Freiburghaus} {et~al.}(1999){Freiburghaus}, {Rosswog}, \&
  {Thielemann}}]{Freiburghaus1999}
{Freiburghaus}, C., {Rosswog}, S., \& {Thielemann}, F.~K. 1999, \apjl, 525,
  L121, \dodoi{10.1086/312343}

\bibitem[{{Fujibayashi} {et~al.}(2018){Fujibayashi}, {Kiuchi}, {Nishimura},
  {Sekiguchi}, \& {Shibata}}]{Fujibayashi2018}
{Fujibayashi}, S., {Kiuchi}, K., {Nishimura}, N., {Sekiguchi}, Y., \&
  {Shibata}, M. 2018, \apj, 860, 64, \dodoi{10.3847/1538-4357/aabafd}

\bibitem[{{Gaigalas} {et~al.}(2019){Gaigalas}, {Kato}, {Rynkun},
  {Rad{\v{z}}i{\={u}}t{\.{e}}}, \& {Tanaka}}]{gaigalas19}
{Gaigalas}, G., {Kato}, D., {Rynkun}, P., {Rad{\v{z}}i{\={u}}t{\.{e}}}, L., \&
  {Tanaka}, M. 2019, \apjs, 240, 29, \dodoi{10.3847/1538-4365/aaf9b8}

\bibitem[{{Goriely} {et~al.}(2015){Goriely}, {Bauswein}, {Just}, {Pllumbi}, \&
  {Janka}}]{Goriely2015}
{Goriely}, S., {Bauswein}, A., {Just}, O., {Pllumbi}, E., \& {Janka}, H.~T.
  2015, \mnras, 452, 3894, \dodoi{10.1093/mnras/stv1526}

\bibitem[{{Gottlieb} \& {Loeb}(2020)}]{Gottlieb2020}
{Gottlieb}, O., \& {Loeb}, A. 2020, \mnras, 493, 1753,
  \dodoi{10.1093/mnras/staa363}

\bibitem[{{Hallinan} {et~al.}(2017){Hallinan}, {Corsi}, {Mooley}, {Hotokezaka},
  {Nakar}, {Kasliwal}, {Kaplan}, {Frail}, {Myers}, {Murphy}, {De}, {Dobie},
  {Allison}, {Bannister}, {Bhalerao}, {Chandra}, {Clarke}, {Giacintucci}, {Ho},
  {Horesh}, {Kassim}, {Kulkarni}, {Lenc}, {Lockman}, {Lynch}, {Nichols},
  {Nissanke}, {Palliyaguru}, {Peters}, {Piran}, {Rana}, {Sadler}, \&
  {Singer}}]{Hallinan2017}
{Hallinan}, G., {Corsi}, A., {Mooley}, K.~P., {et~al.} 2017, Science, 358,
  1579, \dodoi{10.1126/science.aap9855}

\bibitem[{{Hotokezaka} {et~al.}(2013){Hotokezaka}, {Kiuchi}, {Kyutoku},
  {Muranushi}, {Sekiguchi}, {Shibata}, \& {Taniguchi}}]{Hotokezaka2013a}
{Hotokezaka}, K., {Kiuchi}, K., {Kyutoku}, K., {et~al.} 2013, \prd, 88, 044026,
  \dodoi{10.1103/PhysRevD.88.044026}

\bibitem[{{Hotokezaka} \& {Nakar}(2020)}]{Hotokezaka2020}
{Hotokezaka}, K., \& {Nakar}, E. 2020, \apj, 891, 152,
  \dodoi{10.3847/1538-4357/ab6a98}

\bibitem[{{Karp} {et~al.}(1977){Karp}, {Lasher}, {Chan}, \&
  {Salpeter}}]{Karp1977}
{Karp}, A.~H., {Lasher}, G., {Chan}, K.~L., \& {Salpeter}, E.~E. 1977, \apj,
  214, 161, \dodoi{10.1086/155241}

\bibitem[{{Kasen} {et~al.}(2013){Kasen}, {Badnell}, \& {Barnes}}]{kasen2013}
{Kasen}, D., {Badnell}, N.~R., \& {Barnes}, J. 2013, \apj, 774, 25,
  \dodoi{10.1088/0004-637X/774/1/25}

\bibitem[{{Kasen} {et~al.}(2017){Kasen}, {Metzger}, {Barnes}, {Quataert}, \&
  {Ramirez-Ruiz}}]{kasen2017}
{Kasen}, D., {Metzger}, B., {Barnes}, J., {Quataert}, E., \& {Ramirez-Ruiz}, E.
  2017, \nat, 551, 80, \dodoi{10.1038/nature24453}

\bibitem[{{Kasliwal} {et~al.}(2017){Kasliwal}, {Nakar}, {Singer}, \&
  et~al.}]{mansi2017}
{Kasliwal}, M.~M., {Nakar}, E., {Singer}, L.~P., \& et~al. 2017, Science, 358,
  1559, \dodoi{10.1126/science.aap9455}

\bibitem[{{Kawaguchi} {et~al.}(2018){Kawaguchi}, {Shibata}, \&
  {Tanaka}}]{Kawaguchi2018}
{Kawaguchi}, K., {Shibata}, M., \& {Tanaka}, M. 2018, ApJ letter, 865, L21.
\newblock \doarXiv{1806.04088}

\bibitem[{{Korobkin} {et~al.}(2012){Korobkin}, {Rosswog}, {Arcones}, \&
  {Winteler}}]{Korobkin2012}
{Korobkin}, O., {Rosswog}, S., {Arcones}, A., \& {Winteler}, C. 2012, \mnras,
  426, 1940, \dodoi{10.1111/j.1365-2966.2012.21859.x}

\bibitem[{{Kramida} {et~al.}(2018){Kramida}, {Ralchenko}, {Reader}, \& {NIST
  ASD Team}}]{kramida18}
{Kramida}, A., {Ralchenko}, Y., {Reader}, J., \& {NIST ASD Team}. 2018, NIST
  Atomic Spectra Database (version 5.6.1), {\tt https://physics.nist.gov/asd}.
  National Institute of Standards and Technology, Gaithersburg, MD.

\bibitem[{{Kulkarni}(2005)}]{Kulkarni2005}
{Kulkarni}, S.~R. 2005, arXiv e-prints, astro.
\newblock \doarXiv{astro-ph/0510256}

\bibitem[{{Lattimer} \& {Schramm}(1974)}]{Lattimer1974}
{Lattimer}, J.~M., \& {Schramm}, D.~N. 1974, \apjl, 192, L145,
  \dodoi{10.1086/181612}

\bibitem[{{Li} \& {Paczy{\'n}ski}(1998)}]{Li1998}
{Li}, L.-X., \& {Paczy{\'n}ski}, B. 1998, \apjl, 507, L59,
  \dodoi{10.1086/311680}

\bibitem[{Li {et~al.}(2018)Li, Liu, Yu, \& Zhang}]{Li2018}
Li, S.-Z., Liu, L.-D., Yu, Y.-W., \& Zhang, B. 2018, The Astrophysical Journal,
  861, L12, \dodoi{10.3847/2041-8213/aace61}

\bibitem[{{Lippuner} {et~al.}(2017){Lippuner}, {Fern{\'a}ndez}, {Roberts},
  {Foucart}, {Kasen}, {Metzger}, \& {Ott}}]{Lippuner2017}
{Lippuner}, J., {Fern{\'a}ndez}, R., {Roberts}, L.~F., {et~al.} 2017, \mnras,
  472, 904, \dodoi{10.1093/mnras/stx1987}

\bibitem[{{Martin} {et~al.}(2018){Martin}, {Perego}, {Kastaun}, \&
  {Arcones}}]{Martin2018}
{Martin}, D., {Perego}, A., {Kastaun}, W., \& {Arcones}, A. 2018, Classical and
  Quantum Gravity, 35, 034001, \dodoi{10.1088/1361-6382/aa9f5a}

\bibitem[{{Matsumoto} {et~al.}(2018){Matsumoto}, {Ioka}, {Kisaka}, \&
  {Nakar}}]{Matsumoto2018}
{Matsumoto}, T., {Ioka}, K., {Kisaka}, S., \& {Nakar}, E. 2018, \apj, 861, 55,
  \dodoi{10.3847/1538-4357/aac4a8}

\bibitem[{{Metzger} {et~al.}(2015){Metzger}, {Bauswein}, {Goriely}, \&
  {Kasen}}]{Metzger2015}
{Metzger}, B.~D., {Bauswein}, A., {Goriely}, S., \& {Kasen}, D. 2015, \mnras,
  446, 1115, \dodoi{10.1093/mnras/stu2225}

\bibitem[{{Metzger} \& {Fern{\'a}ndez}(2014)}]{Metzger2014}
{Metzger}, B.~D., \& {Fern{\'a}ndez}, R. 2014, \mnras, 441, 3444,
  \dodoi{10.1093/mnras/stu802}

\bibitem[{{Metzger} {et~al.}(2008){Metzger}, {Piro}, \&
  {Quataert}}]{Metzger2008}
{Metzger}, B.~D., {Piro}, A.~L., \& {Quataert}, E. 2008, \mnras, 390, 781,
  \dodoi{10.1111/j.1365-2966.2008.13789.x}

\bibitem[{{Metzger} {et~al.}(2018){Metzger}, {Thompson}, \&
  {Quataert}}]{Metzger2018}
{Metzger}, B.~D., {Thompson}, T.~A., \& {Quataert}, E. 2018, \apj, 856, 101,
  \dodoi{10.3847/1538-4357/aab095}

\bibitem[{{Metzger} {et~al.}(2010){Metzger}, {Mart{\'\i}nez-Pinedo}, {Darbha},
  {Quataert}, {Arcones}, {Kasen}, {Thomas}, {Nugent}, {Panov}, \&
  {Zinner}}]{Metzger2010}
{Metzger}, B.~D., {Mart{\'\i}nez-Pinedo}, G., {Darbha}, S., {et~al.} 2010,
  \mnras, 406, 2650, \dodoi{10.1111/j.1365-2966.2010.16864.x}

\bibitem[{{Mooley} \& {Mooley}(2017)}]{Mooley2017}
{Mooley}, K.~P., \& {Mooley}, S. 2017, GRB Coordinates Network, 22211, 1

\bibitem[{{Perego} {et~al.}(2017){Perego}, {Radice}, \&
  {Bernuzzi}}]{Perego2017}
{Perego}, A., {Radice}, D., \& {Bernuzzi}, S. 2017, \apjl, 850, L37,
  \dodoi{10.3847/2041-8213/aa9ab9}

\bibitem[{{Perego} {et~al.}(2014){Perego}, {Rosswog}, {Cabez{\'o}n},
  {Korobkin}, {K{\"a}ppeli}, {Arcones}, \& {Liebend{\"o}rfer}}]{Perego2014}
{Perego}, A., {Rosswog}, S., {Cabez{\'o}n}, R.~M., {et~al.} 2014, \mnras, 443,
  3134, \dodoi{10.1093/mnras/stu1352}

\bibitem[{{Pinto} \& {Eastman}(2000)}]{Pinto2000}
{Pinto}, P.~A., \& {Eastman}, R.~G. 2000, arXiv e-prints, astro.
\newblock \doarXiv{astro-ph/0006171}

\bibitem[{{Piro} \& {Kollmeier}(2018)}]{Piro2018}
{Piro}, A.~L., \& {Kollmeier}, J.~A. 2018, \apj, 855, 103,
  \dodoi{10.3847/1538-4357/aaaab3}

\bibitem[{{Radice} {et~al.}(2018){Radice}, {Perego}, {Hotokezaka}, {Fromm},
  {Bernuzzi}, \& {Roberts}}]{Radice2018}
{Radice}, D., {Perego}, A., {Hotokezaka}, K., {et~al.} 2018, \apj, 869, 130,
  \dodoi{10.3847/1538-4357/aaf054}

\bibitem[{{Rad{\v{z}}i{\={u}}t{\.{e}}}
  {et~al.}(2020){Rad{\v{z}}i{\={u}}t{\.{e}}}, {Gaigalas}, {Kato}, {Rynkun}, \&
  {Tanaka}}]{radziute20}
{Rad{\v{z}}i{\={u}}t{\.{e}}}, L., {Gaigalas}, G., {Kato}, D., {Rynkun}, P., \&
  {Tanaka}, M. 2020, arXiv e-prints, arXiv:2002.08075.
\newblock \doarXiv{2002.08075}

\bibitem[{{Roberts} {et~al.}(2011){Roberts}, {Kasen}, {Lee}, \&
  {Ramirez-Ruiz}}]{Roberts2011}
{Roberts}, L.~F., {Kasen}, D., {Lee}, W.~H., \& {Ramirez-Ruiz}, E. 2011, \apjl,
  736, L21, \dodoi{10.1088/2041-8205/736/1/L21}

\bibitem[{{Roming} {et~al.}(2005){Roming}, {Kennedy}, {Mason}, {Nousek}, {Ahr},
  {Bingham}, {Broos}, {Carter}, {Hancock}, {Huckle}, {Hunsberger}, {Kawakami},
  {Killough}, {Koch}, {McLelland}, {Smith}, {Smith}, {Soto}, {Boyd},
  {Breeveld}, {Holland}, {Ivanushkina}, {Pryzby}, {Still}, \&
  {Stock}}]{Roming2005}
{Roming}, P. W.~A., {Kennedy}, T.~E., {Mason}, K.~O., {et~al.} 2005, \ssr, 120,
  95, \dodoi{10.1007/s11214-005-5095-4}

\bibitem[{{Rosswog} {et~al.}(2018){Rosswog}, {Sollerman}, {Feindt}, {Goobar},
  {Korobkin}, {Wollaeger}, {Fremling}, \& {Kasliwal}}]{Rosswog2018}
{Rosswog}, S., {Sollerman}, J., {Feindt}, U., {et~al.} 2018, \aap, 615, A132,
  \dodoi{10.1051/0004-6361/201732117}

\bibitem[{{Rybicki} \& {Lightman}(1986)}]{Rybicki1986}
{Rybicki}, G.~B., \& {Lightman}, A.~P. 1986, {Radiative Processes in
  Astrophysics}

\bibitem[{{Sagiv} {et~al.}(2014){Sagiv}, {Gal-Yam}, {Ofek}, {Waxman},
  {Aharonson}, {Kulkarni}, {Nakar}, {Maoz}, {Trakhtenbrot}, {Phinney}, {Topaz},
  {Beichman}, {Murthy}, \& {Worden}}]{Sagiv2014}
{Sagiv}, I., {Gal-Yam}, A., {Ofek}, E.~O., {et~al.} 2014, \aj, 147, 79,
  \dodoi{10.1088/0004-6256/147/4/79}

\bibitem[{{Savchenko} {et~al.}(2017){Savchenko}, {Ferrigno}, {Kuulkers},
  {Bazzano}, {Bozzo}, {Brandt}, {Chenevez}, {Diehl}, {Domingo}, {Hanlon},
  {Jourdain}, {Laurent}, {Lebrun}, {Lutovinov}, {Martin-Carillo}, {Mereghetti},
  {Natalucci}, {Rodi}, {Sunyaev}, \& {Ubertini}}]{Savchenko2017}
{Savchenko}, V., {Ferrigno}, C., {Kuulkers}, E., {et~al.} 2017, in Proceedings
  of the 7th International Fermi Symposium, 58

\bibitem[{{Sekiguchi} {et~al.}(2015){Sekiguchi}, {Kiuchi}, {Kyutoku}, \&
  {Shibata}}]{Sekiguchi2015}
{Sekiguchi}, Y., {Kiuchi}, K., {Kyutoku}, K., \& {Shibata}, M. 2015, \prd, 91,
  064059, \dodoi{10.1103/PhysRevD.91.064059}

\bibitem[{{Sekiguchi} {et~al.}(2016){Sekiguchi}, {Kiuchi}, {Kyutoku},
  {Shibata}, \& {Taniguchi}}]{Sekiguchi2016}
{Sekiguchi}, Y., {Kiuchi}, K., {Kyutoku}, K., {Shibata}, M., \& {Taniguchi}, K.
  2016, \prd, 93, 124046, \dodoi{10.1103/PhysRevD.93.124046}

\bibitem[{{Shibata} {et~al.}(2017){Shibata}, {Fujibayashi}, {Hotokezaka},
  {Kiuchi}, {Kyutoku}, {Sekiguchi}, \& {Tanaka}}]{Shibata2017}
{Shibata}, M., {Fujibayashi}, S., {Hotokezaka}, K., {et~al.} 2017, \prd, 96,
  123012, \dodoi{10.1103/PhysRevD.96.123012}

\bibitem[{{Siegel} \& {Metzger}(2017)}]{Siegel2017}
{Siegel}, D.~M., \& {Metzger}, B.~D. 2017, \prl, 119, 231102,
  \dodoi{10.1103/PhysRevLett.119.231102}

\bibitem[{Tanaka \& Hotokezaka(2013)}]{tanaka2013}
Tanaka, M., \& Hotokezaka, K. 2013, ApJ, 775, 113

\bibitem[{{Tanaka} {et~al.}(2014){Tanaka}, {Hotokezaka}, {Kyutoku}, \&
  et~al.}]{tanaka2014}
{Tanaka}, M., {Hotokezaka}, K., {Kyutoku}, K., \& et~al. 2014, ApJ, 780, 31.
\newblock \doarXiv{1310.2774}

\bibitem[{{Tanaka} {et~al.}(2018){Tanaka}, {Kato}, {Gaigalas}, \&
  et~al.}]{tanaka2018}
{Tanaka}, M., {Kato}, D., {Gaigalas}, G., \& et~al. 2018, ApJ, 852, 109.
\newblock \doarXiv{1708.09101}

\bibitem[{{Tanaka} {et~al.}(2020){Tanaka}, {Kato}, {Gaigalas}, \&
  {Kawaguchi}}]{Tanaka2020a}
{Tanaka}, M., {Kato}, D., {Gaigalas}, G., \& {Kawaguchi}, K. 2020, \mnras, 496,
  1369, \dodoi{10.1093/mnras/staa1576}

\bibitem[{{Tanaka} {et~al.}(2017){Tanaka}, {Utsumi}, {Mazzali}, \&
  et~al.}]{tanaka2017}
{Tanaka}, M., {Utsumi}, Y., {Mazzali}, P.~A., \& et~al. 2017, PASJ, 69, 102.
\newblock \doarXiv{1710.05850}

\bibitem[{{Troja} {et~al.}(2017){Troja}, {Piro}, {van Eerten}, {Wollaeger},
  {Im}, {Fox}, {Butler}, {Cenko}, {Sakamoto}, {Fryer}, {Ricci}, {Lien}, {Ryan},
  {Korobkin}, {Lee}, {Burgess}, {Lee}, {Watson}, {Choi}, {Covino}, {D'Avanzo},
  {Fontes}, {Gonz{\'a}lez}, {Khandrika}, {Kim}, {Kim}, {Lee}, {Lee}, {Kutyrev},
  {Lim}, {S{\'a}nchez-Ram{\'\i}rez}, {Veilleux}, {Wieringa}, \&
  {Yoon}}]{Troja2017}
{Troja}, E., {Piro}, L., {van Eerten}, H., {et~al.} 2017, \nat, 551, 71,
  \dodoi{10.1038/nature24290}

\bibitem[{{Valenti} {et~al.}(2017){Valenti}, {Sand}, {Yang}, {Cappellaro},
  {Tartaglia}, {Corsi}, {Jha}, {Reichart}, {Haislip}, \&
  {Kouprianov}}]{Valenti2017}
{Valenti}, S., {Sand}, D.~J., {Yang}, S., {et~al.} 2017, \apjl, 848, L24,
  \dodoi{10.3847/2041-8213/aa8edf}

\bibitem[{{Verner} {et~al.}(1996){Verner}, {Ferland}, {Korista}, \&
  {Yakovlev}}]{Verner1996}
{Verner}, D.~A., {Ferland}, G.~J., {Korista}, K.~T., \& {Yakovlev}, D.~G. 1996,
  \apj, 465, 487, \dodoi{10.1086/177435}

\bibitem[{{Villar} {et~al.}(2017){Villar}, {Guillochon}, {Berger}, {Metzger},
  {Cowperthwaite}, {Nicholl}, {Alexand er}, {Blanchard}, {Chornock},
  {Eftekhari}, {Fong}, {Margutti}, \& {Williams}}]{Villar2017}
{Villar}, V.~A., {Guillochon}, J., {Berger}, E., {et~al.} 2017, \apjl, 851,
  L21, \dodoi{10.3847/2041-8213/aa9c84}

\bibitem[{{Wanajo} {et~al.}(2014){Wanajo}, {Sekiguchi}, {Nishimura}, \&
  et~al.}]{Wanajo2014}
{Wanajo}, S., {Sekiguchi}, Y., {Nishimura}, N., \& et~al. 2014, \apjl, 789,
  L39, \dodoi{10.1088/2041-8205/789/2/L39}

\bibitem[{{Waxman} {et~al.}(2018){Waxman}, {Ofek}, {Kushnir}, \&
  {Gal-Yam}}]{Waxman2018}
{Waxman}, E., {Ofek}, E.~O., {Kushnir}, D., \& {Gal-Yam}, A. 2018, \mnras, 481,
  3423, \dodoi{10.1093/mnras/sty2441}

\bibitem[{{Wollaeger} {et~al.}(2017){Wollaeger}, {Hungerford}, {Fryer},
  {Wollaber}, {van Rossum}, \& {Even}}]{Wollaeger2017}
{Wollaeger}, R.~T., {Hungerford}, A.~L., {Fryer}, C.~L., {et~al.} 2017, \apj,
  845, 168, \dodoi{10.3847/1538-4357/aa82bd}

\bibitem[{{Wollaeger} {et~al.}(2019){Wollaeger}, {Fryer}, {Fontes}, {Lippuner},
  {Vestrand}, {Mumpower}, {Korobkin}, {Hungerford}, \& {Even}}]{Wollaeger2019}
{Wollaeger}, R.~T., {Fryer}, C.~L., {Fontes}, C.~J., {et~al.} 2019, \apj, 880,
  22, \dodoi{10.3847/1538-4357/ab25f5}

\bibitem[{{Yang} {et~al.}(2017){Yang}, {Valenti}, {Cappellaro}, {Sand },
  {Tartaglia}, {Corsi}, {Reichart}, {Haislip}, \& {Kouprianov}}]{Yang2017}
{Yang}, S., {Valenti}, S., {Cappellaro}, E., {et~al.} 2017, \apjl, 851, L48,
  \dodoi{10.3847/2041-8213/aaa07d}

\bibitem[{{Yu} {et~al.}(2013){Yu}, {Zhang}, \& {Gao}}]{Yu2013}
{Yu}, Y.-W., {Zhang}, B., \& {Gao}, H. 2013, \apjl, 776, L40,
  \dodoi{10.1088/2041-8205/776/2/L40}

\end{thebibliography}
%\bibstyle{ApJ}
\bibliographystyle{aasjournal}

\end{document}